\documentclass[english,twocolumn,transaction,10pt]{IEEEtran}
\usepackage[T1]{fontenc}
\usepackage{float}
\usepackage{amsthm}
\usepackage{amssymb}
\usepackage{stackrel}
\usepackage{graphicx}
\usepackage{setspace}
\usepackage{url}
\usepackage{color}
\usepackage{lipsum} %

\usepackage{caption}
\usepackage{subfigure}
\usepackage{stfloats}

\usepackage[ruled,linesnumbered]{algorithm2e}

\makeatletter

\theoremstyle{plain}

\theoremstyle{definition}

\theoremstyle{plain}

\theoremstyle{plain}

\IEEEoverridecommandlockouts
\usepackage{ifpdf}
\usepackage{cite}
\ifCLASSOPTIONcompsoc
\usepackage[caption=false,font=normalsize,labelfont=sf,textfont=sf]{subfig}
\else
\usepackage[caption=false,font=footnotesize]{subfig}
\fi
\usepackage{amsmath}
\usepackage{mathtools}
\usepackage{booktabs}
\usepackage{multirow}
\usepackage{setspace}
\usepackage{lettrine} %
\usepackage{bm}

\@ifundefined{showcaptionsetup}{}{%
 \PassOptionsToPackage{caption=false}{subfig}}
\usepackage{subfig}
\makeatother

\usepackage{babel}
\newtheorem{define}{Definition}

\newtheorem{lemm}{Lemma}

\newcommand*{\QEDA}{\hfill\ensuremath{\blacksquare}} 

\begin{document}
\captionsetup[figure]{name={Fig.}}

\title{Two-Timescale Optimization for Intelligent Reflecting Surface-Assisted MIMO Transmission in Fast-Changing Channels}
\author{Yashuai Cao,~\IEEEmembership{Student Member,~IEEE}, Tiejun Lv, \emph{Senior Member, IEEE}, \\and Wei Ni, \emph{Senior Member, IEEE}

\thanks{
Manuscript received December 20, 2021; revised April 28, 2022; accepted June 14, 2022. \emph{(Corresponding author: Tiejun Lv)}.

Y. Cao and T. Lv are with the School of Information and Communication Engineering, Beijing University of Posts and Telecommunications (BUPT), Beijing 100876, China (e-mail: \{yashcao, lvtiejun\}@bupt.edu.cn).

W. Ni is with Data61, Commonwealth Scientific and Industrial Research Organisation,
Sydney, NSW 2122, Australia (e-mail: wei.ni@data61.csiro.au).
}}

\maketitle
\begin{abstract}
The application of intelligent reflecting surface (IRS) depends on the knowledge of channel state information (CSI), and has been hindered by the heavy overhead of channel training, estimation, and feedback in fast-changing channels.
This paper presents a new two-timescale beamforming approach to maximizing the average achievable rate (AAR) of IRS-assisted MIMO systems, where the IRS is configured relatively infrequently based on statistical CSI (S-CSI) and the base station precoder and power allocation are updated frequently based on quickly outdated instantaneous CSI (I-CSI).
The key idea is that we first reveal the optimal small-timescale power allocation based on outdated I-CSI yields a water-filling structure.
Given the optimal power allocation, a new mini-batch sampling (mbs)-based particle swarm optimization (PSO) algorithm is developed to optimize the large-timescale IRS configuration with reduced channel samples.
Another important aspect is that we develop a model-driven PSO algorithm to optimize the IRS configuration, which maximizes a lower bound of the AAR by only using the S-CSI and eliminates the need of channel samples.
The model-driven PSO serves as a dependable lower bound for the mbs-PSO.
Simulations corroborate the superiority of the new two-timescale beamforming strategy to its alternatives in terms of the AAR and efficiency, with the benefits of the IRS demonstrated.
\end{abstract}


\section{Introduction}
Recent years have witnessed an unprecedentedly surging demand for mobile data traffic. The demand is expected to continuously grow in the upcoming sixth-generation (6G) cellular systems, resulting from new applications and services \cite{9040264}.
Intelligent reflecting surface (IRS) has been identified as an appealing complementary technology to latest multiple-input multiple-output (MIMO) techniques, to improve spectral and energy efficiency and increase data rates~\cite{8910627, 9140329}. An IRS is a two-dimensional meta-surface array, which can induce favorable scattering to create anomalous reflection and programmable wireless channels via adjusting the phase-shifts of the IRS \cite{9326394}.
Since IRSs are primarily made up of low-cost passive meta-atoms without active radio frequency (RF) chains, they can be densely deployed along with  existing MIMO systems to substantially enhance the spectral efficiency in a cost-effective manner \cite{9110912, 8981888}.
It was reported in \cite{8811733} that the required transmit power of a base station (BS) can decrease approximately quadratically with the increasing number of IRS meta-atoms.
It was also proved that IRSs can achieve better energy efficiency than existing relay systems \cite{mustaghfirin2021, 8888223}.

\subsection{Motivation}
One pressing challenge arising from the incorporation of IRSs into MIMO systems is to overcome the adverse effect of practically imperfect, typically outdated, channel state information (CSI), especially under time-division duplexing (TDD) settings.
Most existing resource allocation algorithms designed for IRS-assisted MIMO systems require the perfect instantaneous CSI (I-CSI), or they can hardly reap the promising gains of the IRSs.
To obtain high-precision CSI, cascaded channel estimation has been widely studied to estimate IRS-enhanced cascaded channels. Many significant progresses, e.g., \cite{9087848, 9361077, 9130088, 9133156}, have been achieved, which makes it possible to obtain nearly perfect I-CSI at MIMO receivers.
The perfect I-CSI is important to effectively design the BS precoder and configure the IRS, but hardly available to the transmitter due to a non-negligible feedback delay, especially in TDD systems \cite{6936326, 8926512, 9120717}.
To circumvent this impasse, some studies have focused on the robust transmissions under imperfect CSI \cite{9270033, 9110587, 9180053, 9473585}.
Two canonical CSI error models, namely, bounded error and statistical error models, were adopted to devise robust beamforming.
In \cite{9374975}, it was reported that configuring the IRS reflection amplitudes is more cost-effective than the phase-shifts under imperfect CSI.
In \cite{8746155}, the passive beamforming of an IRS was optimized based on the upper bound of the ergodic spectral efficiency, where only the statistical CSI (S-CSI) was utilized.

Another challenge arising lies in non-negligible overhead in the phase-shift configurations of IRSs.
In general, the phase-shift configurations of the IRSs are closely coupled with MIMO precoders in IRS-assisted MIMO systems. A holistic, joint design of the two is imperative.
A common approach is to invoke the classic alternating optimization or block coordinate descent \cite{9367432} to obtain a suboptimal solution. Therein, large-scale non-convex unit modulus constraints induced by the physical property of passive meta-atoms need to be dealt with to design the IRS passive beamforming.
Apart from convex relaxation methods, such as semidefinite relaxation \cite{8811733}, alternating direction method of multipliers \cite{9013288}, and manifold optimization \cite{9024490}, a machine learning-based approach was proposed in \cite{9348009} to solve the joint passive beamforming design and resource allocation problem in IRS-aided wireless systems.
Some recent works have applied the particle swarm optimization (PSO) algorithm to the design of IRS passive beamforming, due to its demonstrated effectiveness in solving large-scale nonconvex problems. The authors of \cite{9091558} designed a PSO-based IRS configuration algorithm to reduce the transmit power with no need for the CSI. The algorithm searches for solutions only based on the power and rate metrics of a given channel, and is not suitable for fast-changing channels. An IRS-aided visible light communication system was investigated in \cite{9500409}, where the secrecy rate of the system was maximized using the PSO.

To reduce training overhead, two-timescale beamforming methods~\cite{9198125, 9066923, wang2021massive} have been increasingly studied for IRS.
In \cite{9198125}, active and passive beamforming were optimized at two timescales for an IRS-assisted multiple-input-single-output system. The IRS phase shifts were optimized based on the S-CSI on a frame basis. The BS precoding was optimized based on the I-CSI on a slot basis.
Nadeem~\emph{et al.}~\cite{9066923} studied IRS-aided coverage enhancement where direct BS-user links are unavailable. The IRS phase shifts were asymptotically optimized based on the S-CSI knowledge. The minimum mean square error precoder was designed by using the I-CSI.
By utilizing the channel hardening property, Wang~\emph{et al.}~\cite{wang2021massive} developed a two-timescale protocol, where the large-timescale passive beamforming was optimized to maximize the asymptotically achievable rate of a small-timescale maximal-ratio combining precoder.
Most of these existing IRS configuration algorithms were developed under the assumption that the I-CSI is available and the IRSs are instantaneously reconfigurable. In practice, the acquisition of high-precision CSI is a computationally intensive task.
Let alone the aforementioned feedback delay of the I-CSI.
However, the perfect I-CSI can be hardly available in practice. Only outdated channel samples can be acquired as a result of signaling and feedback overhead \cite{8470152}.

In the face of the challenges of outdated CSI and non-negligible IRS configuration overhead in fast-changing channels, it is of practical interest to design an IRS-assisted MIMO system, where the IRS is configured less frequently but predictively to cater to the fast-changing channel conditions and subsequently frequent BS precoding. The BS precoding needs to tolerate rapidly outdated CSI. To the best of our knowledge, no existing studies have to date considered fast-changing channel conditions in IRS-assisted MIMO systems. Let alone using the outdated CSI in the systems.

\subsection{Contribution}
This paper maximizes the average achievable rate (AAR) of an IRS-assisted, downlink MIMO system under fast-changing channels resulting from mobility and yielding a first-order autoregressive process. This problem is new and has never been addressed in the literature, because the acquired CSI can be quickly and substantially outdated when used for precoding in the fast-changing channel.
The key contributions of the paper are summarized as follows.
\begin{itemize}
\item
We design a new two-timescale beamforming strategy: the IRS is configured at a large timescale on a frame basis, while the transmit and receive beamforming (i.e., precoder and detection) is performed at a small timescale on a slot basis. The inherent temporal correlation of the channel is exploited between consecutive time slots to create effective parallel data streams, with the inter-stream interference resulting from the outdated CSI suppressed at the user equipment (UE).
\item
We reveal that the optimal small-timescale transmit power allocation only depends on the outdated CSI and can be readily obtained by water-filling techniques.
\item
With the proposed two-timescale beamforming strategy and the optimal small-timescale power allocation in the fast-changing channel, we design a new data-driven PSO algorithm with a mini-batch recursive sampling surrogate function as the fitness function, called mbs-PSO, which divides the random channel samples into batches and expedites evaluating the fitness of the IRS passive beamforming by sampling the batches.
\item
We derive an analytical lower bound of the AAR, and maximize the lower bound by solving a trace-of-inverse-covariance minimization (TicMin) problem. The new objective of TicMin depends directly on the S-CSI, and is efficiently solved by designing a new model-driven PSO, called LBO-PSO, which evaluates the trace-of-inverse-covariance for the fitness. With no need for channel samples, the LBO-PSO is a computationally efficient alternative to the mbs-PSO algorithm.
\end{itemize}
Comprehensive simulations verify the superiority of the proposed two-timescale beamforming strategy to five benchmarks in terms of the AAR and efficiency. The benefits of the IRS are clearly demonstrated with useful insights shed.

\subsection{Organization and Notation}
The remainder of this paper is outlined as follows. In Section \ref{sec:syst}, we describe the channel model, system architecture, and formulate the new two-timescale beamforming strategy. In Section \ref{sec:method}, we optimize the precoding and power allocation for the new two-timescale strategy. Water-filling-based power allocation is developed at a small timescale. PSO-based IRS configuration are designed at a large timescale, to maximize the AAR of the system. Simulation results are provided in Section \ref{sec:sim}, followed by conclusions in Section \ref{sec:con}.

\emph{Notations}: Lower- and upper-case boldface letters indicate column-vectors and matrices, re-spectively; $(\cdot)^{\mathsf{T}}$ and $(\cdot)^{\mathsf{H}}$ denote transpose and conjugate transpose, respectively; $(\cdot)^{-1}$ and $(\cdot)^{\dagger}$ stand for inverse and pseudo-inverse, respectively; $\mathrm{trace}(\cdot)$ returns the trace of a matrix; $\mathrm{angle}\{\cdot\}$ returns the angle of a complex value; $\mathrm{diag}\{\cdot\}$ returns a diagonal matrix; $\mathbb{E}\{\cdot\}$ denotes expectation; $\otimes$ is the Kronecker product operator; $[\cdot]_{(m,n)}$ denotes the $(m,n)$-th element of a matrix; $\mathbb{C}^{m \times n}$ is the space of $m\times n$ complex matrices; and $\mathbf{I}_{n}$ is the $n \times n$ identity matrix.

\section{System Setting and Problem Formulation}\label{sec:syst}
In this section, we first describe the IRS-assisted MIMO system model and the channel model with an emphasis on the outdated CSI.
The MIMO system allows the BS to produce its precoder based on the outdated CSI. The UE generates its detector by using a zero-forcing (ZF) criterion and I-CSI, suppressing the inter-stream interference arising from the outdated CSI.
We put forth a new two-timescale passive beamforming and power allocation problem with the problem statement provided at last.

\subsection{System Model and Channel Model}
\begin{figure}[t]
	\centering{}\includegraphics[scale=0.5]{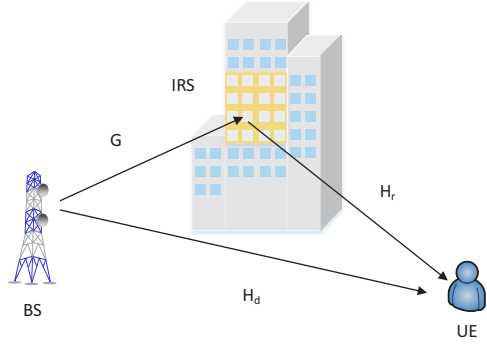}
	\caption{An illustration of the considered IRS-assisted downlink point-to-point MIMO system, where both the BS and UE are equipped with multiple antennas, and the channels are fast-changing, e.g., due to the movement of the UE.}
	\label{fig:fig1}
\end{figure}

As illustrated in Fig. 1, we consider an IRS-assisted MIMO link from a BS to a UE under a TDD setting.
The BS is equipped with $N_{\mathrm{t}}$ transmit antennas. The UE is equipped with $N_{\mathrm{r}}$ receive antennas.
The IRS is a uniform square planar array (USPA) consisting of $N=N_x N_y$ meta-atoms, with $N_x$ meta-atoms per row and $N_y$ meta-atoms per column.
As shown in Fig. \ref{fig:fig1}, $\mathbf{H}_{\mathrm{d}}$ denotes the direct channel from the BS to the UE, and $\mathbf{G}$ and $\mathbf{H}_{\mathrm{r}}$ denote the BS-IRS and IRS-UE channels, respectively. For notational brevity, we define $\mathcal{H}=\{\mathbf{G}, \mathbf{H}_{\mathrm{d}}, \mathbf{H}_{\mathrm{r}}\}$ as the full channel ensemble.

The IRS is deployed to achieve the line-of-sight (LoS) propagation between the BS and the IRS. Typically, there are few scatters around the BS and the IRS, but rich scatters around the UE. We assume that the BS-IRS channel is \emph{rank-one}, and the IRS-UE channel is a multi-path channel. The geometry of the IRS is known to the BS, and so is the BS-IRS channel. The BS-IRS channel is assumed to be time-invariant, due to the stationarity of both
the BS and IRS.
The BS-IRS channel is given by
\begin{align}
\mathbf{G}= \sqrt{L_{\mathrm{br}}} \mathbf{a}(\phi) \mathbf{b}(\phi_{\mathrm{e}}, \phi_{\mathrm{a}})^{\mathsf{H}} \in \mathbb{C}^{N_{\mathrm{t}} \times N},
\end{align}
where $L_{\mathrm{br}}$ is large-scale fading, $\phi$ is the angle-of-departure (AoD) at the BS, $\phi_{\mathrm{e}}(\text{or} \ \phi_{\mathrm{a}})$ is the elevation (or azimuth) angles-of-arrival (AoA) at the IRS, $\mathbf{a}(\phi)$ is the array steering vector of the BS, and $\mathbf{b}(\phi_{\mathrm{e}}, \phi_{\mathrm{a}})$ is the array steering vector of the IRS. The array steering vectors are
\begin{align}
\mathbf{a}(\phi) &= [1, \cdots, e^{-j2\pi(N_t-1)\frac{d}{\lambda}\sin\phi}]^{\mathsf{T}} \in \mathbb{C}^{N_t \times 1},\\
\mathbf{b}(\phi_{\mathrm{e}}, \phi_{\mathrm{a}}) &= [1, \cdots, e^{-j2\pi \frac{d \left((N_y-1)\sin\phi_{\mathrm{e}}\sin\phi_{\mathrm{a}} + (N_x-1)\cos\phi_{\mathrm{e}} \right) }{\lambda} }]^{\mathsf{T}} \nonumber\\
&= [1,\cdots, e^{-j2\pi(N_y-1)\frac{d}{\lambda}\sin\phi_{\mathrm{e}} \sin\phi_{\mathrm{a}} }]^{\mathsf{T}} \otimes \nonumber\\
&\quad\quad [1,\cdots, e^{-j2\pi (N_x-1)\frac{d}{\lambda}\cos\phi_{\mathrm{e}} }]^{\mathsf{T}} \nonumber\\
&=\mathbf{b}_{N_y}(\phi_{\mathrm{e}}, \phi_{\mathrm{a}}) \otimes \mathbf{b}_{N_x}(\phi_{\mathrm{e}}) \in \mathbb{C}^{N\times 1},
\end{align}
where $\lambda$ is the carrier wavelength, and $d$ is the spacing between adjacent antenna elements of the BS or adjacent passive meta-atoms of the IRS.

We focus on a time frame, within which the S-CSI of all links remains unchanged. The time frame is divided evenly into $T$ time slots. The channel coefficients of the links, i.e., I-CSI, do not change during a time slot. The non-LoS (NLoS) channel coefficient matrices are correlated across different time slots.
At time slot $t$, the IRS-UE channel can be given
\begin{align}
\mathbf{H}_{\mathrm{r}} [t] = \sqrt{L_{\mathrm{ur}}} \left(  \kappa \bar{\mathbf{H}}_{\mathrm{r}} + \sqrt{1-\kappa^2}\tilde{\mathbf{H}}_{\mathrm{r}} [t] \right)\in \mathbb{C}^{N \times N_{\mathrm{r}}},
\end{align}
where $L_{\mathrm{ur}}$ accounts for large-scale fading; $\kappa$ is the Rician fading-related factor; $\bar{\mathbf{H}}_{\mathrm{r}}$ and $\tilde{\mathbf{H}}_{\mathrm{r}}\sim \mathcal{CN}(0,\mathbf{I})$ are the deterministic LoS component and the NLoS component, respectively.
Likewise, the BS-UE channel can be given by
\begin{align}
\mathbf{H}_{\mathrm{d}}[t] = \sqrt{L_{\mathrm{bu}}} \left( \kappa \bar{\mathbf{H}}_{\mathrm{d}} + \sqrt{1-\kappa^2} \tilde{\mathbf{H}}_{\mathrm{d}} [t] \right)\in \mathbb{C}^{N_{\mathrm{t}} \times N_{\mathrm{r}}}.
\end{align}
where $L_{\mathrm{bu}}$, $\bar{\mathbf{H}}_{\mathrm{d}}$ and $\tilde{\mathbf{H}}_{\mathrm{d}}$ are defined in the same way as $L_{\mathrm{ur}}$, $\bar{\mathbf{H}}_{\mathrm{r}}$ and $\tilde{\mathbf{H}}_{\mathrm{r}}$, respectively.

The temporal evolution of the NLoS Rayleigh fading channel is modeled by the first-order autoregressive process \cite{6937178}, as given by
\begin{align}
\tilde{\mathbf{H}}_{\mathrm{r}} [t] = \rho\tilde{\mathbf{H}}_{\mathrm{r}} [t-1] + \mathbf{E}[t], \
\tilde{\mathbf{H}}_{\mathrm{d}} [t] = \rho\tilde{\mathbf{H}}_{\mathrm{d}} [t-1] + \mathbf{E}[t],\label{eq:delay}
\end{align}
where $\mathbf{E}[t] \sim \mathcal{CN}\left(0, (1-\rho^2 )\mathbf{I} \right)$ yields the circular complex Gaussian distribution. Following the Jakes' model \cite{8753608}, the time correlation coefficient is $\rho=J_0(2\pi f_d \tau)=J_0(2\pi\bar{f}_d)$. Here, $J_0(\cdot)$ is the zero-th order Bessel function of the first kind; $f_d$ is the maximum Doppler frequency; $\tau$ is the delay between the time when the CSI is measured and the time when the BS produces the precoders based on the CSI; and $\bar{f}_d = f_d \tau$ is the normalized Doppler frequency.

It is reasonable to assume that the channel ensemble $\mathcal{H}$ can be accurately estimated.
In a TDD system with channel reciprocity assumed, the BS can estimate the CSI based on the pilots sent by the UE in the uplink and then use the CSI to precode the subsequent downlink transmissions. $\tau$ can be up to a slot. In time-varying channels, the CSI estimated in the uplink can be outdated when used for downlink transmissions.
On the other hand, the channel reciprocity may not hold in a practical TDD system because the RF chains are generally different/asymmetric in the uplink and downlink.
The UE needs to estimate the CSI in the downlink based on the pilots sent by the BS, and then feeds the CSI back to the BS in the uplink, as done in FDD systems, as specified in 3GPP standards~\cite{etsi136lte}. $\tau$ is no shorter than a slot.
Given the non-negligible delay $\tau$, the channels may have changed substantially between when they are measured and when they are used for transmissions.

When considering both channel estimation errors and channel aging, the channel vector can be rewritten as $\tilde{\mathbf{H}}_{\mathrm{r}} [t] = \rho'\tilde{\mathbf{H}}_{\mathrm{r}}^{\mathrm{est}} [t-1]  + \mathbf{E} [t] + \mathbf{E}_{\mathrm{est}}[t]$, where $\tilde{\mathbf{H}}_{\mathrm{r}}^{\mathrm{est}}$ denotes the estimated channel, $\mathbf{E} \sim \mathcal{CN}\left(\mathbf{0}, (1-\rho'^2)(1-\sigma_e^2) \mathbf{I} \right)$ accounts for the channel aging, $\mathbf{E}_{\mathrm{est}} \sim \mathcal{CN}\left(\mathbf{0}, (1-\rho'^2)\sigma_e^2\mathbf{I} \right)$ accounts for the channel estimation error, $\rho'$ is a channel correlation coefficient, and $\sigma_e^2$ is the estimation error variance.	
With the statistical independence between $\mathbf{E}_{\mathrm{est}}$ and $\mathbf{E}$, we can rewrite the channel vector as
$\tilde{\mathbf{H}}_{\mathrm{r}} [t] = \rho' \tilde{\mathbf{H}}_{\mathrm{r}}^{\mathrm{est}} [t-1] + \mathbf{E}' [t]$
where $\mathbf{E}' [t] = \mathbf{E}[t] + \mathbf{E}_{\mathrm{est}}[t] \sim \mathcal{CN} (\mathbf{0}, (1-\rho'^2 ) \mathbf{I})$. The channel aging and channel estimation error can be aggregated as one estimation error that exhibits strong analogy to (\ref{eq:delay}), and can be dealt with in the same way as (\ref{eq:delay}).

\subsection{System Architecture}
In the considered, IRS-assisted, $N_{\mathrm{t}} \times N_{\mathrm{r}}$ MIMO system, the BS transmit $M (M \le N_{\mathrm{t}})$ data streams to the UE.
At the $t$-th slot, the received signal at the UE is given by
\begin{align}
\mathbf{y}[t]
&= \mathbf{W}_{\mathrm{r}}^{\mathsf{H}} [t]   \underbrace{ \left( \mathbf{H}_{\mathrm{d}}^{\mathsf{H}} [t] + \mathbf{H}_{\mathrm{r}}^{\mathsf{H}}[t] \mathbf{\Theta}^{\mathsf{H}}[t]  \mathbf{G}^{\mathsf{H}} [t] \right) }_{\text{effective channel}} \mathbf{W}_{\mathrm{t}}[t] \nonumber\\ &\mathbf{\Lambda}^{\frac{1}{2}} [t] \mathbf{x}[t]  + \mathbf{n}[t]    \nonumber \\
&= \mathbf{W}_{\mathrm{r}}^{\mathsf{H}} [t] \check{\mathbf{H}}^{\mathsf{H}} [t]  \mathbf{W}_{\mathrm{t}} [t] \mathbf{\Lambda}^{\frac{1}{2}} [t]   \mathbf{x}[t]  + \mathbf{n}[t] ,
\end{align}
where $\mathbf{x} \in \mathbb{C}^{M \times 1}$ is the transmit signal, $\mathbf{n} \sim \mathcal{CN}(0, \sigma_n^2 \mathbf{I}_{N_{\mathrm{r}}})$ is the additive Gaussian noise with $\mathbb{E}( {\mathbf{n}}  {\mathbf{n}}^{\mathsf H})=\sigma_n^2 \mathbf{I}_{M}$, $\mathbf{W}_{\mathrm{t}}$ is the precoding matrix at the BS, $\mathbf{W}_{\mathrm{r}}$ is the detection matrix at the UE, $\mathbf{\Lambda}=\mathrm{diag}\left\{[P_1, \cdots, P_{M}]\right\} \in \mathbb{C}^{M \times M}$ collects the transmit powers allocated to each data stream, $\mathbf{\Theta} =\mathrm{diag} \left\{\exp (j\boldsymbol{\theta}) \right\} \in \mathbb{C}^{N\times N}$ is the diagonal reflection matrix of the IRS, and $\check{\mathbf{H}} \triangleq \mathbf{H}_{\mathrm{d}}  + \mathbf{G} \mathbf{\Theta} \mathbf{H}_{\mathrm{r}} \in \mathbb{C}^{N_{\mathrm{t}} \times N_{\mathrm{r}}}$ is defined as \emph{effective channel} for notational brevity.

\begin{figure}[t]
	\centering{}\includegraphics[scale=0.35]{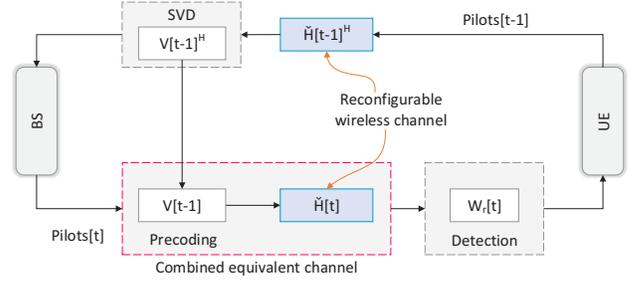}
	\caption{The considered transmission architecture for fast-changing channels, where the BS precodes the data symbols and pilots based on the truncated SVD of outdated CSI, and then the UE estimates the I-CSI based on the precoded pilots and detects data signals.}
	\label{fig:fig2}
\end{figure}

In the time-varying channels, the precoding vector produced at the BS based on the outdated CSI can lead to non-negligible inter-stream interference, deteriorating the AAR performance. For this reason, we adopt an SVD-ZF beamforming approach~\cite{6399078} and extend it under power constraints to optimize the power allocation based on the outdated CSI, as illustrated in Fig. \ref{fig:fig2}.		
Specifically, the BS produces the precoder $\mathbf{W}_{\mathrm{t}}[t]=\mathbf{V}[t-1]$ based on the outdated CSI available at the BS, where $\mathbf{V}[t-1]$ is the right unitary matrix of the singular value decomposition (SVD) of $\check{\mathbf{H}}[t-1]$. Both the pilot signals and data symbols are precoded and transmitted through the \emph{equivalent end-to-end channel}, $\underline{\mathbf{H}}[t] \triangleq \mathbf{V}[t-1]^{\mathsf{H}} \check{\mathbf{H}}[t] \in \mathbb{C}^{M \times N_{\mathrm{r}}}$. The UE can estimate $\underline{\mathbf{H}}[t]$ based on the precoded pilots.		
The estimation is straightforward, and can be readily done on a slot basis, as typically done in conventional wireless communication systems~\cite{7174567, 8885990, 6200884}.

Based on the estimated equivalent channel, i.e., I-CSI, the UE detects the data symbols using the following zero-forcing (ZF) detection matrix:
\begin{align}
	\mathbf{W}_{\mathrm{r}}[t]  & \triangleq \Big( \check{\mathbf{H}}^{\mathsf{H}}[t] \mathbf{V}[t-1] \Big)^{\dagger} \nonumber\\
	&= \left[ \Big(  \check{\mathbf{H}}^{\mathsf{H}}[t] \mathbf{V}[t-1] \Big)^{\mathsf{H}}  \check{\mathbf{H}}^{\mathsf{H}}[t] \mathbf{V}[t-1]   \right]^{-1} \nonumber \\
	&\Big(  \check{\mathbf{H}}^{\mathsf{H}} [t]\mathbf{V}[t-1] \Big)^{\mathsf{H}} \nonumber \\
	&= \left( \underline{\mathbf{H}}[t] \underline{\mathbf{H}}^{\mathsf{H}}[t] \right)^{-1}  \underline{\mathbf{H}}[t] .
\end{align}
The received signal is obtained as
\begin{align}
	\mathbf{y}[t] &= \mathbf{W}_{\mathrm{r}}^{\mathsf{H}}[t] \left( \check{\mathbf{H}}^{\mathsf{H}}[t] \mathbf{V}[t-1] \mathbf{\Lambda}^{\frac{1}{2}} [t] \mathbf{x} [t]+ \mathbf{n}[t] \right) \nonumber\\
	&= \left( \underline{\mathbf{H}}[t] \underline{\mathbf{H}}^{\mathsf{H}}[t] \right)^{-1}  \underline{\mathbf{H}}[t] \underline{\mathbf{H}}^{\mathsf{H}}[t] \mathbf{\Lambda}^{\frac{1}{2}}[t]  \mathbf{x}[t] + \nonumber \\
	&\left( \underline{\mathbf{H}}[t] \underline{\mathbf{H}}^{\mathsf{H}}[t] \right)^{-1}  \underline{\mathbf{H}}[t]  \mathbf{n}[t] \nonumber \\
	&=  \mathbf{\Lambda}^{\frac{1}{2}} [t] \mathbf{x}[t] + \left( \underline{\mathbf{H}}[t] \underline{\mathbf{H}}^{\mathsf{H}}[t] \right)^{-1}  \underline{\mathbf{H}}[t]  \mathbf{n} [t].
\end{align}
The signal-to-interference-plus-noise ratio (SINR) of the $m$-th data stream is given by
\begin{align}
	\gamma_m[t] &= \frac{\Big[\mathbf{\Lambda}[t] \Big]_{(m,m)}}{\sigma_n^2 \left[ \left( ( \check{\mathbf{H}}^{\mathsf{H}}[t] \mathbf{V}[t-1] )^{\mathsf{H}} (  \check{\mathbf{H}}^{\mathsf{H}}[t] \mathbf{V}[t-1] ) \right)^{-1} \right]_{(m,m)} } \nonumber\\
	&=\frac{ P_m [t] }{\sigma_n^2 \left[ \left( \underline{\mathbf{H}}[t] \underline{\mathbf{H}}^{\mathsf{H}}[t] \right)^{-1}  \right]_{(m,m)} }.
\end{align}

This beamforming strategy can efficiently eliminate the inter-stream interference resulting from the SVD precoding based on the outdated CSI.
The SVD precoding with water-filling power allocation is also known to be optimal under I-CSI~\cite{6399078},~\cite{eldad2008next}. It is still likely to produce effective parallel streams based on the outdated CSI, by taking advantage of the temporal correlation of the autoregressive channels.
To this end, the small-timescale SVD-ZF beamforming strategy is increasingly closer to the optimum, as the channel changes more slowly and undergoes stronger temporal correlation. The strategy becomes optimal when the channel stops changing. On the other hand, it may not be plausible to jointly optimize the precoders and detectors in the considered system, due to the fast-changing channels and subsequently outdated CSI at the BS. Since the BS is the one that optimizes the precoders, by no means can it come up with the optimal detector that requires the I-CSI.

\subsection{Beamforming Strategy and Problem Formulation}
It is not straightforward to estimate the effective channel $\check{\mathbf{H}}$, because part of it can be augmented by the IRS. The passive IRS meta-atoms are made up by PIN-diodes, and do not involve active RF chains or any signal processing. For this reason, cascaded channel estimation is widely adopted to estimate the effective channels~\cite{9133156}.
The cascaded channel estimation needs to be conducted, every time the IRS is reconfigured.
In consequence, unacceptably large training overhead and delay would be required, if the reflection phase-shifts of the IRS are reconfigured excessively frequently. Moreover, it is non-trivial to synchronize the BS/UE and the IRS if the IRS is reconfigured as frequently as the BS in fast-changing channels. Guard intervals need to be reserved for the reconfigurations at the cost of spectrum utilization.

\begin{figure}[t]
	\centering{}\includegraphics[scale=0.4]{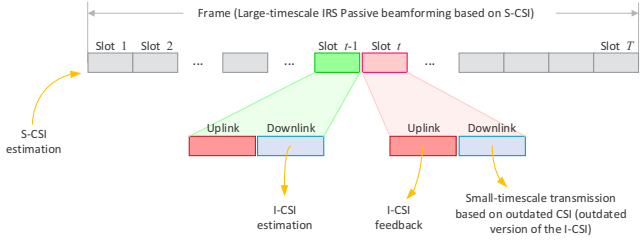}
	\caption{Pictorial illustration of the new two-timescale beamforming strategy, where the S-CSI is estimated once per frame and used for the large-timescale passive beamforming of the IRS. In contrast, the power allocation of the BS is updated slot by slot, based on the outdated CSI available to the BS, due to the non-negligible delay between the I-CSI estimation (e.g., at the UE) and the power allocation.}
	\label{fig:fig3}
\end{figure}

We propose a new two-timescale beamforming strategy, which configures the IRS once per frame based on the statistical knowledge of the CSI and updates the BS precoders every time slot based on available outdated CSI.
As illustrated in Fig. \ref{fig:fig3}, the large-scale fading coefficients $\{ L_{\mathrm{ur}}, L_{\mathrm{bu}} \}$ and the S-CSI ensemble $\{\bar{\mathbf{H}}_{\mathrm{r}}, \bar{\mathbf{H}}_{\mathrm{d}}\}$ are evaluated at the beginning of a frame based on all CSI estimated in the preceding frames, using the S-CSI estimation approach~\cite{9369969, 9408385}, and assumed to remain unchanged within the frame. Based on the S-CSI, the BS optimizes the IRS passive beamforming for the entire frame.
Given the large-timescale IRS passive beamforming for a frame, the small-timescale precoding and detection can be conducted slot by slot throughout the frame, as described in Section II-B. Specifically, the BS performs the SVD precoding based on the outdated CSI, and the UE conducts the ZF detector based on the I-CSI.

The goal of the new two-timescale beamforming strategy is to maximize the AAR of the considered IRS-assisted MIMO system by jointly optimizing the long-term passive beamforming vector of the IRS and the per-slot power allocation of the BS.
This problem is formulated as
\begin{subequations}
\begin{align}
\text{(P1)}: \quad & \underset{ \mathbf{\Theta} }{\max} \  \mathbb{E} \left\{ \underset{ \mathbf{\Lambda} }{\max} \  \sum_{m=1}^{M}
\log \Big[ 1+ \gamma_m (\mathbf{\Theta}, \mathbf{\Lambda}[t]) \Big]  \right\} \nonumber \\
\mathrm{s.t.} \quad &   \sum_{m=1}^{M} P_m  = P_{\mathrm{tot}},  \quad P_m \ge 0 , \forall m, \label{eq:power} \\
& \left\vert [\mathbf{\Theta}]_{n,n} \right\vert = 1, \quad  \mathrm{angle}\left( [\mathbf{\Theta}]_{n,n} \right) \in [0, 2\pi), \forall n. \label{eq:theta}
\end{align}
\end{subequations}
where (\ref{eq:power}) guarantees the total transmit power of the BS does not exceed its power budget $P_{\mathrm{tot}}$, and (\ref{eq:theta}) specifies the unit-modulus nature and the phase-shift range of the IRS meta-atoms.

\section{Proposed Two-timescale Beamforming Strategy}\label{sec:method}
In this section, we reformulate Problem (P1) as a joint design of power allocation at the BS and passive beamforming at the IRS, and design new modified PSO algorithms to solve the problem efficiently.
We first rewrite the SINR of the $m$-th data stream at slot $t$, $\gamma_m(P_m[t], \mathbf{\Theta}, \mathcal{H}[t-1])$, as
\begin{align}
& \gamma_m(P_m[t], \mathbf{\Theta}, \mathcal{H}[t-1]) \nonumber\\
=& \frac{ P_m[t] }{\sigma_n^2 \left[ \left\{ \mathbf{V}^{\mathsf{H}}[t-1]  \check{\mathbf{H}}[t] \check{\mathbf{H}}[t]^{\mathsf{H}} \mathbf{V}[t-1] \right\}^{-1} \right]_{(m, m)}} \nonumber\\
=& \frac{ P_m[t] }{\sigma_n^2 \left[ \left\{ \underline{\mathbf{H}}[t] \underline{\mathbf{H}}[t]^{\mathsf{H}} \right\}^{-1} \right]_{(m, m)} } \nonumber\\
=& \frac{ P_m[t] }{\sigma_n^2 \left[ \mathbf{\Phi}^{-1}[t] \right]_{(m, m)} }
= \frac{ P_m[t] }{\sigma_n^2 f_m( \boldsymbol{\theta} )[t] } , \label{eq:sinr}
\end{align}
where $\mathbf{\Phi}[t] \triangleq \underline{\mathbf{H}}[t]\underline{\mathbf{H}}^{\mathsf{H}}[t] \in \mathbb{C}^{M \times M}$ is the covariance matrix of the equivalent channel, and $f_m( \boldsymbol{\theta} )[t]=\left[\mathbf{\Phi}^{-1}(\boldsymbol{\theta} ) [t] \right]_{(m,m)}$ is the $m$-th diagonal element of the inverse of the covariance matrix.
Problem (P1) can be rewritten as
\begin{subequations}
\begin{align}
\text{(P1')}: \quad & \underset{ \boldsymbol{\theta} }{\max} \  \mathbb{E} \left\{ \underset{ \mathbf{p} }{\max} \  \sum_{m=1}^{M}
\log \left[ 1+ \frac{P_m [t]}{\sigma_n^2 f_m( \boldsymbol{\theta} ) [t]} \right]  \right\} \nonumber \\
\mathrm{s.t.} \quad &  \mathbf{p} \succeq 0, \quad \mathbf{1}^{\mathsf{T}}\mathbf{p} = P_{\mathrm{tot}}, \label{eq:power1} \\
&  \mathbf{\Theta}= \mathrm{diag}\left\{ e^{ j\boldsymbol{\theta} } \right\},  {\theta}_{n} \in [0, 2\pi), \forall n,  \label{eq:theta1}
\end{align}
\end{subequations}
where $\mathbf{p}=[P_1, \cdots, P_M]^{\mathsf{T}}$ collects the transmit powers of the $M$ data streams.

Problem (P1') is a stochastic optimization problem, because of the per-slot optimization of the power allocation at the small timescale.
At each time slot, the power allocation is based on an outdated cascaded channel measurement, and the fixed phase-shifts of the IRS configured at the beginning of the current frame.
On the other hand, the optimization of the IRS configuration needs to predictively maximize the expectation of the AAR over random channel samples and potential per-slot power allocations throughout the frame.
The typical approach which transforms a stochastic optimization problem into a deterministic non-convex optimization problem, would confront serious challenges, i.e., 1) lack of explicit relation between $\gamma_i$ and $\boldsymbol{\theta}$; 2) nonconvex unit-modulus constraints on $\mathbf{\Theta}$; and 3) the need of minimizing each diagonal entry of the inverse of the covariance matrix, i.e., $\left[ \mathbf{\Phi}^{-1}\right]_{(m,m)}$, which is arduous.

\subsection{Small-timescale Power Allocation}
Provided that the IRS passive beamforming is optimized at the beginning of the frame; at the $t$-th slot of the frame, with the outdated CSI, $\check{\mathbf{H}}[t-1]$, the original Problem (P1) reduces to:
\begin{align}
	\text{(P2)}: \quad
	\underset{ \mathbf{p} }{\max} \  \sum_{m=1}^{M}
	\log \left( 1+ \frac{P_m[t] f_m^{-1}(\boldsymbol{\theta}) [t] }{\sigma_n^2 } \right) \quad
	\mathrm{s.t.} \    \text{(\ref{eq:power1})}  \nonumber.
\end{align}
Problem (P2) is convex with respect to the variable $\mathbf{p}$, and can be solved using the off-the-peg water-filling algorithm~\cite{tse2005fundamentals}. With $\mu$ being the Lagrange multiplier for constraint (\ref{eq:power1}) and $[x]^{+}= \max\{x, 0\}$, the optimal per-slot power allocation is given by~\cite{tse2005fundamentals}
\begin{align}
	P_m^{\mathrm{opt}} [t]=  \left[ \frac{1}{\mu \ln 2}- \sigma_n^2 f_m(\boldsymbol{\theta}) [t] \right]^{+}. \label{eq:opt_pow}
\end{align}
The corresponding per-slot sum-rate is given by
\begin{align}
	&\sum_{m=1}^{M}
	\log \left( 1+ \frac{P_m [t]}{\sigma_n^2 f_m(\boldsymbol{\theta})[t] } \right) \nonumber\\
	=& \sum_{m=1}^{M}
	\log \left( 1+ \frac{1}{\sigma_n^2} \left[ \frac{ 1 }{ \mu f_m(\boldsymbol{\theta}) [t] \ln 2 } - \sigma_n^2\right]^{+} \right).
\end{align}

The proposed framework is still operational even if the channel is independent between different time slots. In this case, the SVD precoding produces a transmit beamformer independent of the current channel, and can be viewed as an arbitrary precoding matrix for the current slot. Given the ZF detector at the UE, the inter-stream interference can be suppressed. Nonetheless, the use of an arbitrary precoding receive no benefit from the transmit power allocation relying on effective decomposition of the channel into parallel streams.

\subsection{Data-driven Large-timescale Passive Beamforming with Mini-batch-sampling-based PSO}
The per-frame passive beamforming is a stochastic optimization problem due to the expectation operation in the objective of Problem (P1').
To circumvent this impasse, we first rewrite the objective in a deterministic form.
Specifically, we take the average rate of $L_B$ random channel samples to approximate the expectation in the objective.
Suppose that the $l$-th sample of the IRS-related time-variant channels in the $t$-th time slot is $\hat{\mathcal{H}}_{\mathrm{irs}}^{l}[t]= \{ \hat{\mathbf{H}}_{\mathrm{d}}^{l}[t-1], \hat{\mathbf{H}}_{\mathrm{r}}^{l}[t-1] \}$, accounting for the outdated CSI at the BS. The AAR of the $m$-th data-stream is approximated by
\begin{align}
&\bar{r}_m \left(P_m^{\mathrm{opt}}( \mathbf{\Theta}, \hat{\mathcal{H}}_{\mathrm{irs}} ), \mathbf{\Theta} \right)
\nonumber\\
=& \frac{1}{T} \frac{1}{L_B} \sum_{t=1}^{T} \sum_{l=1}^{L_B} \log \left[ 1 +  \gamma_m \left( P_m[t], \mathbf{\Theta}, \hat{\mathcal{H}}_{\mathrm{irs}}^{l}[t-1] \right)  \right]. \label{eq:AAR_old}
\end{align}

The large-timescale passive beamforming problem is cast as
\begin{subequations}
\begin{align}
\text{(P3)}: \quad   \underset{ \mathbf{\Theta} }{\max} \  \sum_{m=1}^{M} \bar{r}_m \left(P_m^{\mathrm{opt}}( \mathbf{\Theta}, \hat{\mathcal{H}}_{\mathrm{irs}} ), \mathbf{\Theta} \right)  \quad
\mathrm{s.t.} \  \mathrm{(\ref{eq:theta1})}. \nonumber
\end{align}
\end{subequations}
Apart from the nonconvex constant-modulus constraint, the power allocation variables are closely coupled with the reflection phase-shifts of the IRS. Moreover, to maximize the AAR involves minimizing the diagonal entries of the inverse of the covariance matrix, i.e., $\left[\mathbf{\Phi}^{-1}\right]_{(m,m)}$, which makes it difficult to optimize $\mathbf{\Theta}$ since $\left[\mathbf{\Phi}^{-1}\right]_{(m,m)}$ cannot be written as an explicit function of $\mathbf{\Theta}$.
Conventional convex optimization methods or convexification methods, such as successive convex relaxation, cannot be applied to solve Problem (P3).

We propose a new PSO framework to solve Problem (P3).
The PSO, a widely-used meta-heuristic bionic optimization algorithm~\cite{8926512}, can find the optimal solution with little prior information. It enjoys the benefits of fewer parameters, simple calculation implementation, and fast convergence~\cite{jian2020hybrid}.
In the PSO method, the position of a particle stands for a potential solution. The fitness function of a particle is typically defined to be the optimization objective.
We define a passive reflection angle vector $\boldsymbol{\theta}$ as the position of a particle, which corresponds to a passive beamforming matrix $\mathbf{\Theta}=\mathrm{diag}\{e^{j\boldsymbol{\theta}}\}$.
The objective of Problem (P3), i.e., the AAR, is used as the fitness function, as given by
\begin{align}
J(\boldsymbol{\theta}) = \sum_{m=1}^{M} \bar{r}_m \left(P_m^{\mathrm{opt}}( \boldsymbol{\theta}, \hat{\mathcal{H}}_{\mathrm{irs}} ), \boldsymbol{\theta} \right).
\label{eq:fit}
\end{align}
We note that, for each particle $\boldsymbol{\theta}$, the optimal power solution can be uniquely found according to (\ref{eq:opt_pow}). This guarantees the uniqueness of the fitness function in (\ref{eq:fit}).
We assume that $P$ particles are employed to seek for the optimal solutions in the constrained search space and each is assigned with a velocity at every iteration.
The set of particle positions, denoted by $\mathcal{L}_P = \{\boldsymbol{\theta}_1, \boldsymbol{\theta}_2, \cdots, \boldsymbol{\theta}_P\}$, corresponds to the set of passive beamforming matrices $\{\mathbf{\Theta}_1, \mathbf{\Theta}_2, \cdots, \mathbf{\Theta}_P\}$.
In every iteration, a particle is assessed based on its fitness function to determine whether the current position implies a good solution. The particle records its best position ever found.

The global optimal position is selected from the best positions of all particles.
Let $\dot{\boldsymbol{\theta}}_{p}^{(i)}$ and $\ddot{\boldsymbol{\theta}}^{(i)}$ denote the best position of the $p$-th particle and the global optimal position of all particles at the $i$-th iteration, respectively.
The $p$-th particle updates its position according to its velocity $\mathbf{v}_{p}^{(i)}$ at the $i$-th iteration.
The resulting new position is used to update the best position of this particle, if the fitness value of the previous best position is lower than that of the new position. Otherwise, the best position of this particle remains unchanged. After one round of the fitness value evaluation, the global optimal position of all particles is accordingly updated.
Both the velocity and position of the $p$-th particle at the $i$-th iteration are updated by
\begin{align}
\mathbf{v}_{p}^{(i)} &= w \mathbf{v}_{p}^{(i-1)} + c_1 \varepsilon_1 (\dot{\boldsymbol{\theta}}_{p}^{(i-1)} - \boldsymbol{\theta}_{p}^{(i-1)}) + \nonumber\\
&c_2 \varepsilon_2 (\ddot{\boldsymbol{\theta}}^{(i-1)} - \boldsymbol{\theta}_{p}^{(i-1)}), \label{eq:speed}\\
\boldsymbol{\theta}_{p}^{(i)} &= \boldsymbol{\theta}_{p}^{(i-1)} + \mathbf{v}_{p}^{(i)}, \label{eq:pos}
\end{align}
where $w$ is the nonnegative inertia weight of a particle; $c_1$ and $c_2$ denote cognitive and social scaling factors, respectively; and $\varepsilon_1$ and $\varepsilon_2$ are two independent random variables with a uniform distribution in $(0,1)$. In addition, the control parameters, such as the swarm size $P$ and iteration number $I_{\mathrm{iter}}$, are critical for the convergence rate and global search performance.
Once the parameters updated in (\ref{eq:pos}) are outside $(-\pi, \pi]$, their boundary values are taken.

Each particle requires a notable amount of calculations for the fitness function evaluation per iteration, since the fitness function is composed of a large number of AARs over random channel samples.
It can be computationally prohibitive to take large-scale channel samples (which are four-dimensional matrices with size of $N_{\mathrm{t}}\times N_{\mathrm{r}}\times L_B \times P$) to obtain a fitness function value per iteration, potentially hindering the convergence of the PSO model.

We develop a new mini-batch recursive sampling (mbs)-PSO algorithm which is able to substantially reduce the number of channel samples used to evaluate the fitness function while reaping satisfactory performance. Specifically, we introduce a \emph{mini-batch recursive sampling surrogate function} \cite{8470152} to replace the fitness function in (\ref{eq:fit}).
All $L_B$ random channel samples are partitioned into $N_B$ batches with $L_{\mathrm{mb}} = L_B/N_B$ samples per batch.
At each iteration, the mini-batch sampling surrogate function is given by
\begin{align}
&J^{(i)} (\boldsymbol{\theta}) = (1-\mu^{(i)}) J^{(i-1)}   + \mu^{(i)} \sum_{m=1}^{M} \sum_{t=1}^{T} \nonumber\\
& \sum_{l=(i-1)L_{\mathrm{mb}}+1}^{i L_{\mathrm{mb}}}\frac{ \log \left[ 1 +  \gamma_m \left( P_m[t], \boldsymbol{\theta}, \hat{\mathcal{H}}_{\mathrm{irs}}^{l}[t-1] \right)  \right] }{T L_{\mathrm{mb}}}, \label{eq:fitness}
\end{align}
where $\mu^{(i)}$ is an iteration-dependent constant accounting for the decay weight coefficient associated with the new sampling of the AAR.
Here, we set $\mu^{(i)}=i^{-0.2}$ according to the diminishing stepsize rules \cite{7412752}, which is commonly used to guarantee more accurate sampling approximation as $i$ increases in the stochastic optimization.

The proposed mbs-PSO algorithm is summarized in Algorithm \ref{alg:batch_pso}. The algorithm is \emph{data-driven} in spirit, given its use of random channel samples.
Given the S-CSI estimated at the beginning of a frame, the BS can generate the $L_B$ channel samples to approximate the future channels in the frame used for downlink transmissions. Recall that there are $T$ slots in the frame. The BS can generate the NLOS components of $\frac{L_B}{T}$ channel samples for the IRS-UE channel for the $t$-th of the $T$ slots, as given by
\begin{align}
	\tilde{\mathbf{H}}_{\mathrm{r}}^{l}[t] = \rho\tilde{\mathbf{H}}_{\mathrm{r}}^{l}[t-1] + \mathbf{E}[t], \ l=1,\cdots,\frac{L_B}{T}; t=1,\cdots,T,
\end{align}
where $\tilde{\mathbf{H}}_{\mathrm{r}}^{l} \sim \mathcal{CN}(0,\mathbf{I})$ is the NLoS component of the $l$-th IRS-UE channel sample.
Then, the $\frac{L_B}{T}$ channel samples of the IRS-UE channel for the $t$-th slot are given by
\begin{align}
	\mathbf{H}_{\mathrm{r}}^{l} [t] &= \sqrt{L} \left(  \kappa \bar{\mathbf{H}}_{\mathrm{r}} + \sqrt{1-\kappa^2}\tilde{\mathbf{H}}_{\mathrm{r}}^{l} [t] \right), \nonumber\\
	& l=1,\cdots,\frac{L_B}{T}; t=1,\cdots,T,
\end{align}
where $L$ is the large-scale fading coefficient and $\bar{\mathbf{H}}_{\mathrm{r}}$ is a constant LoS component during the frame.
By repeating this process from $t=1$ to $T$, we can obtain $\frac{L_B}{T}$ series of time-varying channel samples to approximate the IRS-UE channels of the upcoming $T$ slots.
The $L_B$ samples of the BS-UE channel, $\mathbf{H}_{\mathrm{d}}^{l} [t]$, can be obtained in the same way as $\mathbf{H}_{\mathrm{r}}^{l} [t]$.

\begin{algorithm}[htbp]
\small
\caption{Proposed mbs-PSO Algorithm}
\label{alg:batch_pso}
\LinesNumbered
  \SetKwBlock{Begin}{Step 1:}{end}
  \Begin(\textbf{Swarm initialization}){
  \SetKwInOut{KwIn}{\textbf{Initialize the PSO parameters}} \KwIn{$\{P, w, c_1, c_2, \varepsilon_1, \varepsilon_2, I_{\mathrm{iter}}\}$.}
  \SetKwInOut{KwIn}{\textbf{Initialize the position and velocity}} \KwIn{$\{\boldsymbol{\theta}_p^{(0)}\}_{p=1}^{P},  \{\mathbf{v}_{p}^{(0)}\}_{p=1}^{P}$.}
  Generate $L_B$ outdated channel samples according to the S-CSI and first-order AR model\;
  Separate the total channel samples into $N_B$ mini batches with each $L_{\mathrm{mb}}$ samples\;
  For each particle $\boldsymbol{\theta}_p^{(0)}$, obtain the associated optimal powers $\mathbf{p}^{(0)}_p$ by using (\ref{eq:opt_pow})\;
  Evaluate the fitness value in (\ref{eq:fitness}) of each particle using first mini-batch of samples, and obtain the initial best position set $\{\dot{\boldsymbol{\theta}}_p\}_{p=1}^{P}$\;
  Find the global optimal position $\ddot{\boldsymbol{\theta}} = \arg\max_{\dot{\boldsymbol{\theta}}} \{ J^{(0)} (\dot{\boldsymbol{\theta}}_1), J^{(0)} (\dot{\boldsymbol{\theta}}_2), \cdots, J^{(0)} (\dot{\boldsymbol{\theta}}_P) \}$\;}
  \SetKwBlock{Begin}{Step 2:}{end}
  \Begin(\textbf{Iterative search}){
  \For{$i=1: I_{\mathrm{iter}}$}{{
  Update particle velocity $\{\mathbf{v}_{p}^{(i)}\}_{p=1}^{P}$ and position $\{\boldsymbol{\theta}_p^{(i)}\}_{p=1}^{P}$ by using (\ref{eq:speed}) and (\ref{eq:pos})\;
  \For{$p=1: P$}{
  Given particle $\boldsymbol{\theta}_p^{(i)}$, calculate the optimal powers $\mathbf{p}^{(i)}_p$ according to (\ref{eq:opt_pow})\;
  Evaluate the fitness value of particle $\boldsymbol{\theta}_p^{(i)}$ over $i^{\mathrm{th}}$ mini-batch of samples by using (\ref{eq:fitness})\;
  \eIf{$J^{(i)} > J^{(i-1)}$}
  {$\dot{\boldsymbol{\theta}}_p = \boldsymbol{\theta}_p^{(i)}$\;}{$\dot{\boldsymbol{\theta}}_p = \boldsymbol{\theta}_p^{(i-1)}$\;}
  Update $\ddot{\boldsymbol{\theta}}_p = \arg\max_{\dot{\boldsymbol{\theta}}} \{ J^{(i)} (\dot{\boldsymbol{\theta}}_1), J^{(i)} (\dot{\boldsymbol{\theta}}_2), \cdots, J^{(i)} (\dot{\boldsymbol{\theta}}_P) \}$\;
  }
  }
  }
  }
\end{algorithm}

\subsection{Model-driven Large-timescale Passive Beamforming with Lower-Bound-Oriented PSO}
We proceed to develop a new \emph{model-driven} PSO method for optimizing the IRS passive beamforming, which only exploits the S-CSI and the known random distributions of the channels. The new model-driven PSO algorithm, referred to as Lower-Bound-Oriented PSO (LBO-PSO), can avoid the use of random channel samples (cf. Algorithm \ref{alg:batch_pso}), thereby reducing its reliance on wireless data samples and the risk of approximation inaccuracy.
Following are the details of the LBO-PSO algorithm, where the slot index $t$ is dropped for illustration convenience.

First, by using the arithmetic-geometric inequality \cite{7166320}, the lower bound for the objective of Problem (P1) is given by (\ref{eq:lower}), shown at the top of the next page.
\begin{figure*}
\begin{align}
&\mathbb{E} \left\{\sum_{m=1}^{M}
\log \left( 1+ \frac{P_m}{\sigma_n^2 \left[  \mathbf{\Phi}^{-1} \right]_{(m,m)} } \right)  \right\}
\ge   \mathbb{E} \left\{ M
\log \left( 1 + \frac{M}{\sigma_n^2 \sum_{m=1}^{M} P_m^{-1}\left[  \mathbf{\Phi}^{-1} \right]_{(m,m)} } \right)  \right\}  \nonumber \\
\ge & M \log \left( 1+ \frac{M}{\sigma_n^2 \mathbb{E} \left\{ \sum_{m=1}^{M} P_m^{-1}\left[ \mathbf{\Phi}^{-1} \right]_{(m,m)} \right\} } \right)
= M \log \left( 1+ \frac{\tilde{M}}{ \sum_{m=1}^{M} P_m^{-1} \mathbb{E} \left\{ \left[ \mathbf{\Phi}^{-1}  \right]_{(m,m)} \right\} } \right) \nonumber \\
=& M \log \left( 1 + \frac{\tilde{M}}{ \mathrm{trace} \left( \mathbf{\Lambda}^{-1} \mathbb{E} \left\{  \mathbf{\Phi}^{-1} \right\} \right) } \right), \label{eq:lower}
\end{align}
\hrulefill
\end{figure*}
In (\ref{eq:lower}), $\tilde{M} \triangleq M/\sigma_n^2$, $\mathbf{\Lambda}^{-1}=\mathrm{diag}\left\{ [P_1^{-1}, \cdots, P_M^{-1}]\right\}$, and the second inequality is based on Jensen's inequality.
Instead of directly solving Problem (P1'), we maximize the lower bound of Problem (P1') here, as given by
\begin{align}
\text{(P4)}: \quad &\underset{ \boldsymbol{\theta}, \mathbf{p} }{\max} \ \log \left( 1+ \frac{\tilde{M}}{   \mathrm{trace} \left( \mathbf{\Lambda}^{-1} \mathbb{E} \left\{  \mathbf{\Phi}^{-1} (\boldsymbol{\theta})   \right\} \right) } \right)  \nonumber\\
&\mathrm{s.t.}  \  \mathrm{(\ref{eq:power1})}, \mathrm{(\ref{eq:theta1})}, \nonumber
\end{align}
which can be further equivalently rewritten as
\begin{align}
\text{(P5)}:  \quad \underset{ \boldsymbol{\theta}, \mathbf{p} }{\min} \  \mathrm{trace} \left( \mathbf{\Lambda}^{-1} \mathbb{E} \left\{   \mathbf{\Phi}^{-1}(\boldsymbol{\theta}) \right\} \right)\quad
\mathrm{s.t.} \  \mathrm{(\ref{eq:power1})}, \mathrm{(\ref{eq:theta1})}. \nonumber
\end{align}
Next, we derive an analytical approximation of the objective in Problem (P5) by invoking the property of the \emph{complex inverse Wishart matrix} \cite{8501915}.
We prove that $\mathbf{\Phi}$ yields the Wishart distribution by first proving that $\check{\mathbf{H}}$ is a complex Gaussian distributed matrix. Without loss of generality, we assume that $\check{\mathbf{H}} \in \mathbb{C}^{N_{\mathrm{t}} \times N_{\mathrm{r}}}$.
We expand matrix $\check{\mathbf{H}} \in \mathbb{C}^{N_{\mathrm{t}} \times N_{\mathrm{r}}}$ as
\begin{align}
	\check{\mathbf{H}} =& \mathbf{H}_{\mathrm{d}}  + \mathbf{G} \mathbf{\Theta} \mathbf{H}_{\mathrm{r}} \nonumber \\
	=& \sqrt{L_{\mathrm{bu}}} \left( \kappa \bar{\mathbf{H}}_{\mathrm{d}} + \sqrt{1-\kappa^2} \tilde{\mathbf{H}}_{\mathrm{d}} \right) + \nonumber\\
	\quad & \mathbf{G} \mathbf{\Theta} \left( \sqrt{L_{\mathrm{ur}}} \left(  \kappa \bar{\mathbf{H}}_{\mathrm{r}} + \sqrt{1-\kappa^2}\tilde{\mathbf{H}}_{\mathrm{r}} \right) \right) \nonumber \\
	=& \sqrt{\kappa^2 L_{\mathrm{bu}}} \bar{\mathbf{H}}_{\mathrm{d}} + \sqrt{\kappa^2 L_{\mathrm{ur}}}\mathbf{G} \mathbf{\Theta}\bar{\mathbf{H}}_{\mathrm{r}} + \sqrt{\kappa_{\perp}^2 L_{\mathrm{bu}}} \tilde{\mathbf{H}}_{\mathrm{d}} \nonumber \\
	&+  \sqrt{\kappa_{\perp}^2 L_{\mathrm{ur}} }\mathbf{G} \mathbf{\Theta} \tilde{\mathbf{H}}_{\mathrm{r}},
\end{align}
where $\check{\mathbf{H}}=[\check{\mathbf{h}}_1, \check{\mathbf{h}}_2, \cdots, \check{\mathbf{h}}_{N_{\mathrm{r}}} ]$ and $\kappa_{\perp}^2 = 1-\kappa^2$.

For ease of exposition, we define $\bar{\mathbf{H}}_{\mathrm{d}}=[\bar{\mathbf{d}}_1, \bar{\mathbf{d}}_2, \cdots, \bar{\mathbf{d}}_{N_{\mathrm{r}}}]$, $\tilde{\mathbf{H}}_{\mathrm{d}}=[\tilde{\mathbf{d}}_1, \tilde{\mathbf{d}}_2, \cdots, \tilde{\mathbf{d}}_{N_{\mathrm{r}}}]$, $\bar{\mathbf{H}}_{\mathrm{r}}=[\bar{\mathbf{r}}_1, \bar{\mathbf{r}}_2, \cdots, \bar{\mathbf{r}}_{N_{\mathrm{r}}}]$ and $\tilde{\mathbf{H}}_{\mathrm{r}}=[\tilde{\mathbf{r}}_1, \tilde{\mathbf{r}}_2, \cdots, \tilde{\mathbf{r}}_{N_{\mathrm{r}}}]$.
Since $\tilde{\mathbf{H}}_{\mathrm{d}}$ and $\tilde{\mathbf{H}}_{\mathrm{r}}$ are mutual independent and both contain complex Gaussian variables, we have $\tilde{\mathbf{d}}_{j} \sim \mathcal{CN}(\mathbf{0}, \mathbf{I}_{N_{\mathrm{t}}})$ and $\tilde{\mathbf{r}}_{j} \sim \mathcal{CN}(\mathbf{0}, \mathbf{I}_{N}), \ \forall j=1,2,\cdots, N_{\mathrm{r}}$.
Since the linear transformation of a Gaussian random vector or the sum of independent Gaussian vectors is Gaussian~\cite{9618945}, the $j$-th column of $\check{\mathbf{H}}$ yields
\begin{align}
	\check{\mathbf{h}}_j &= \sqrt{\kappa^2 L_{\mathrm{bu}}} \bar{\mathbf{d}}_j + \sqrt{\kappa^2 L_{\mathrm{ur}}}\mathbf{G} \mathbf{\Theta}\bar{\mathbf{r}}_j + \sqrt{\kappa_{\perp}^2 L_{\mathrm{bu}}} \tilde{\mathbf{d}}_j \nonumber \\
	&+ \sqrt{\kappa_{\perp}^2 L_{\mathrm{ur}} }\mathbf{G} \mathbf{\Theta} \tilde{\mathbf{r}}_j \sim \mathcal{CN} \Big( \sqrt{\kappa^2 L_{\mathrm{bu}}} \bar{\mathbf{d}}_j + \sqrt{\kappa^2 L_{\mathrm{ur}}}\mathbf{G} \mathbf{\Theta}\bar{\mathbf{r}}_j, \nonumber\\
	& \kappa_{\perp}^2 L_{\mathrm{bu}} \mathbf{I}_{N_{\mathrm{t}}} +  \kappa_{\perp}^2 L_{\mathrm{ur}} \mathbf{G} \mathbf{G}^{\mathsf H} \Big).
\end{align}
Define $\underline{\mathbf{H}}=[\underline{\mathbf{h}}_1, \underline{\mathbf{h}}_2, \cdots, \underline{\mathbf{h}}_{N_{\mathrm{r}}}]$.
Likewise, the $j$-th column of $\underline{\mathbf{H}}$ yields $\underline{\mathbf{h}}_j = \mathbf{V}^{\mathsf{H}} \check{\mathbf{h}}_j
\sim \mathcal{CN} ( \sqrt{\kappa^2 L_{\mathrm{bu}}}  \mathbf{V}^{\mathsf{H}}\bar{\mathbf{d}}_j + \sqrt{\kappa^2 L_{\mathrm{ur}}} \mathbf{V}^{\mathsf{H}} \mathbf{G} \mathbf{\Theta}\bar{\mathbf{r}}_j, \kappa_{\perp}^2 L_{\mathrm{bu}} \mathbf{I}_{M} +  \kappa_{\perp}^2 L_{\mathrm{ur}} \mathbf{V}^{\mathsf{H}} \mathbf{G} \mathbf{G}^{\mathsf H} \mathbf{V} )$.
As a result, $\underline{\mathbf{H}}$ is a complex Gaussian distributed matrix, i.e., $\underline{\mathbf{H}} \sim \mathcal{CN} \left( \mathbf{S}_{\underline{\mathbf{H}} } , \mathbf{\Gamma}_{ \underline{\mathbf{H}} } \right)$, where $\mathbf{S}_{\underline{\mathbf{H}} } = \sqrt{\kappa^2 L_{\mathrm{bu}}} \mathbf{V}^{\mathsf{H}}\bar{\mathbf{H}}_{\mathrm{d}} + \sqrt{\kappa^2 L_{\mathrm{ur}}} \mathbf{V}^{\mathsf{H}} \mathbf{G} \mathbf{\Theta}\bar{\mathbf{H}}_{\mathrm{r}}$ and $\mathbf{\Gamma}_{ \underline{\mathbf{H}} } =\mathbf{I}_{N_{\mathrm{r}}} \otimes \left(\kappa_{\perp}^2 L_{\mathrm{bu}} \mathbf{I}_{M} +  \kappa_{\perp}^2 L_{\mathrm{ur}} \mathbf{V}^{\mathsf{H}}\mathbf{G} \mathbf{G}^{\mathsf H} \mathbf{V} \right)$.

\begin{define}
	Suppose a random matrix $\mathbf{X} \in \mathbb{C}^{q \times p}\ (q \le p)$, where any two columns $\mathbf{x}_i$ and $\mathbf{x}_j$ are independent for any $i,j\ (i \neq j)$. Each column corresponds to a covariance matrix ${\mathbf \Sigma}={\mathbb{E}}\left[{\mathbf x}_{i}{\mathbf x}_{i}^{\mathsf{H}}\right],\forall_{i}$.
	If every element of $\mathbf{X}$ follows the complex Gaussian distribution, i.e., $\mathbf{x}_{ij} \sim \mathcal{CN}(0, 1)$ and $\mathbb{E}(\mathbf{X}) = \mathbf{0}$, then the Hermitian matrix ${\mathbf X}{\mathbf X}^{\mathsf H} \sim {\mathcal W}_{q}(p, {\mathbf \Sigma})$ yields the central Wishart distribution.
	For $\mathbb{E}(\mathbf{X}) = \mathbf{M} \neq \mathbf{0}$, ${\mathbf X}{\mathbf X}^{\mathsf H}\sim {\mathcal W}_{q}(p, \mathbf{M}, {\mathbf \Sigma})$ follows the non-central Wishart. 	
	\label{def:re1_5}
\end{define}

According to \textbf{Definition} \ref{def:re1_5}, $\mathbf{\Phi}=\underline{\mathbf{H}} \underline{\mathbf{H}}^{\mathsf{H}}$ follows a non-central Wishart distribution, i.e.,
\begin{equation}
	\mathbf{\Phi} \sim \mathcal{W}_{M} \left( N_{\mathrm{r}}, \mathbf{M}_{\mathbf{\Phi} }, \mathbf{\Sigma}_{\mathbf{\Phi} } \right).
	\label{eq:noncent}
\end{equation}
In (\ref{eq:noncent}), for notational brevity, we define
\begin{align}
	\mathbf{M}_{\mathbf{\Phi} }
	&= \kappa_{\perp}^2 L_{\mathrm{bu}} \mathbf{I}_{M} +  \kappa_{\perp}^2 L_{\mathrm{ur}} \mathbf{V}^{\mathsf{H}}\mathbf{G} \mathbf{G}^{\mathsf H} \mathbf{V}, \\
	\mathbf{\Sigma}_{\mathbf{\Phi} }
	&= \mathbf{M}_{\mathbf{\Phi} }^{-1} \mathbf{S}_{\underline{\mathbf{H}} } \mathbf{S}_{\underline{\mathbf{H}} }^{\mathsf H} = \mathbf{M}_{\mathbf{\Phi} }^{-1}  \mathbf{V}^{\mathsf{H}} \Big( \bar{\mathbf{R}}_{\mathrm{dd}} + \bar{\mathbf{R}}_{\mathrm{dr}} \mathbf{\Theta}^{\mathsf{H}} \mathbf{G}^{\mathsf{H}} + \nonumber\\ &\mathbf{G}\mathbf{\Theta} \bar{\mathbf{R}}_{\mathrm{rd}} + \mathbf{G}\mathbf{\Theta}  \bar{\mathbf{R}}_{\mathrm{rr}}   \mathbf{\Theta}^{\mathsf{H}} \mathbf{G}^{\mathsf{\mathsf{H}}} \Big)\mathbf{V},
\end{align}
where $\bar{\mathbf{R}}_{\mathrm{dd}} = L_{\mathrm{bu}}\kappa^2 \bar{\mathbf{H}}_\mathrm{d} \bar{\mathbf{H}}_{\mathrm{d}}^{\mathsf{H}} \in \mathbb{C}^{N_{\mathrm{t}} \times N_{\mathrm{t}}}$, $\bar{\mathbf{R}}_{\mathrm{dr}} = \kappa \kappa_{\perp} \sqrt{ L_{\mathrm{bu}} L_{\mathrm{ur}} }  \bar{\mathbf{H}}_\mathrm{d} \bar{\mathbf{H}}_\mathrm{r}^{\mathsf{H}} \in \mathbb{C}^{N_{\mathrm{t}} \times N}$, $\bar{\mathbf{R}}_{\mathrm{rd}} = \bar{\mathbf{R}}_{\mathrm{dr}}^{\mathsf{H}}$, and $\bar{\mathbf{R}}_{\mathrm{rr}} = L_{\mathrm{ur}}\kappa^2 \bar{\mathbf{H}}_\mathrm{r} \bar{\mathbf{H}}_\mathrm{r}^{\mathsf{H}} \in \mathbb{C}^{N \times N}$.

\begin{define}
	Suppose that matrix $\mathbf{A}=\mathbf{X}\mathbf{X}^{\mathsf{H}}$ follows the Wishart distribution of $\mathcal{W}_{M_A}(N_A, \mathbf{M}_{A}, \mathbf{\Sigma}_A)$, where $N_A$ is the degree of freedom (with $N_A \ge M_A$), $\mathbf{M}_{A}$ is the mean matrix of $\mathbf{H}$, and $\mathbf{\Sigma}_A \in \mathbb{C}^{M_A \times M_A}$ is the covariance matrix of $\mathbf{H}$. Then, the inverse $\mathbf{A}^{-1}$ has the mean
	\begin{align}
		\mathbb{E}(\mathbf{A}^{-1})=\frac{\mathbf{\Sigma}_A^{-1}}{N_A - M_A}.
	\end{align}
	where $\mathbb{E}(\mathbf{A})= N_A \mathbf{\Sigma}_A$. The mean of the inverse is larger than the inverse of the mean, since $\mathbb{E}(\mathbf{A})^{-1}=\frac{\mathbf{\Sigma}_A^{-1}}{N_A} < \mathbb{E}(\mathbf{A}^{-1})$.
	\label{def:wish}
\end{define}

\begin{IEEEproof}
	The detailed proof can be found in~\cite{6816003}.
\end{IEEEproof}

By approximating (\ref{eq:noncent}) with the central Wishart distribution with the same first-order moment \cite{8501915}, we obtain the central Wishart distribution of $\mathbf{\Phi}$ as follows:
\begin{align}
	\mathbf{\Phi} \sim \mathcal{W}_{M} \left( N_{\mathrm{r}}, \mathbf{M}_{\mathbf{\Phi} } + \frac{1}{N_{\mathrm{r}}} \mathbf{S}_{\underline{\mathbf{H}} } \mathbf{S}_{\underline{\mathbf{H}} }^{\mathsf H} \right).
\end{align}
According to \textbf{Definition} 2, the expectation of the matrix inverse is given by~\cite{9618945}
\begin{align}
	\mathbb{E}\left(\mathbf{\Phi}^{-1} \right)
	= \frac{\left( \mathbf{M}_{\mathbf{\Phi} } + \frac{1}{N_{\mathrm{r}}} \mathbf{S}_{\underline{\mathbf{H}} } \mathbf{S}_{\underline{\mathbf{H}} }^{\mathsf H} \right)^{-1}}{  N_{\mathrm{r}}- M}.
\end{align}

The objective of Problem (P5) can be equivalently rewritten as
\begin{align}
&\mathrm{trace} \left( \mathbf{\Lambda}^{-1} \mathbb{E} \left\{ \mathbf{\Phi}^{-1}(\boldsymbol{\theta}) \right\} \right) \nonumber \\
=& \mathrm{trace} \left( \mathbf{\Lambda}^{-1} \frac{\left( \mathbf{M}_{\mathbf{\Phi} } + \frac{1}{N_{\mathrm{r}}} \mathbf{S}_{\underline{\mathbf{H}} } \mathbf{S}_{\underline{\mathbf{H}} }^{\mathsf H} \right)^{-1}}{N_{\mathrm{r}} - M}  \right) \nonumber\\
=& \frac{\mathrm{trace} \left( \mathbf{\Lambda}^{-1}  \left( \mathbf{M}_{\mathbf{\Phi} } + \frac{1}{N_{\mathrm{r}}} \mathbf{S}_{\underline{\mathbf{H}} } \mathbf{S}_{\underline{\mathbf{H}} }^{\mathsf H} \right)^{-1} \right)}{N_{\mathrm{r}} - M} . \label{eq:low1}
\end{align}

\begin{lemm}
Let $\mathbf{A} \succ \mathbf{0}$ be an $N\times N$ symmetric positive definite matrix and $\mathbf{B} \succeq \mathbf{0}$ be an $N\times N$ positive semidefinite matrix. Then,
\begin{align}
\operatorname{trace}(\mathbf{A} \mathbf{B}) \le &
\Vert \mathbf{B} \Vert_2 \operatorname{trace}(\mathbf{A}) = \lambda_{\max}(\mathbf{B}) \operatorname{trace}(\mathbf{A}) \nonumber\\
\le& \operatorname{trace}(\mathbf{A}) \operatorname{trace}(\mathbf{B}).
\end{align}
Since $\Vert \mathbf{B} \Vert_2 < \operatorname{trace}(\mathbf{B})$, $\lambda_{\max}(\mathbf{B}) \operatorname{trace}(\mathbf{A})$ provides a tighter upper bound than $\operatorname{trace}(\mathbf{A}) \operatorname{trace}(\mathbf{B})$ for $\operatorname{trace}(\mathbf{AB})$. \label{lemm:tr}
\end{lemm}

\begin{IEEEproof}
See Appendix \ref{prf1}.
\end{IEEEproof}

By further applying \emph{Lemma} \ref{lemm:tr}, (\ref{eq:low1}) is upper bounded by
\begin{align}
&\frac{\mathrm{trace} \left( \mathbf{\Lambda}^{-1}  \mathbf{D}^{-2}  \right)}{N_{\mathrm{t}} - M} \nonumber\\
\le &\frac{ \lambda_{\max}( \mathbf{\Lambda}^{-1}) }{N_{\mathrm{t}} - M} \mathrm{trace} \left(  \left( \mathbf{M}_{\mathbf{\Phi} } + \frac{1}{N_{\mathrm{r}}} \mathbf{S}_{\underline{\mathbf{H}} } \mathbf{S}_{\underline{\mathbf{H}} }^{\mathsf H} \right)^{-1} \right) \nonumber \\
= & \frac{ N_{\mathrm{r}} \lambda_{\max}( \mathbf{\Lambda}^{-1}) }{N_{\mathrm{t}} - M} \mathrm{trace}\left[ \left( N_{\mathrm{r}} \mathbf{M}_{\mathbf{\Phi} } + \mathbf{S}_{\underline{\mathbf{H}} } \mathbf{S}_{\underline{\mathbf{H}} }^{\mathsf H} \right)^{-1} \right].
\label{eq:lowercom}
\end{align}

As a result, the objective of Problem (P5) can be updated as
\begin{align}
\text{(P6)}: \ \underset{\boldsymbol{\theta}, \mathbf{p}}{\min} \ \mathrm{trace}\left[ \left( N_{\mathrm{r}} \mathbf{M}_{\mathbf{\Phi} } + \mathbf{S}_{\underline{\mathbf{H}} } \mathbf{S}_{\underline{\mathbf{H}} }^{\mathsf H} \right)^{-1} \right] \
\mathrm{s.t.} \  \mathrm{(\ref{eq:power1})}, \mathrm{(\ref{eq:theta1})}. \nonumber
\end{align}

With reference to the proposed mbs-PSO method in Algorithm \ref{alg:batch_pso}, the proposed LBO-PSO method takes a potential passive reflection angle vector $\boldsymbol{\theta}$ as the position of each particle.
Different from the mbs-PSO method, the fitness function of the LBO-PSO algorithm is defined to be the objective of Problem (P6), as given by
\begin{align}
J_{\mathrm{low}}(\boldsymbol{\theta}) = \mathrm{trace}\left[  \left( N_{\mathrm{r}} \mathbf{M}_{\mathbf{\Phi} } + \mathbf{S}_{\underline{\mathbf{H}} } \mathbf{S}_{\underline{\mathbf{H}} }^{\mathsf H} \right)^{-1}  \right].
\label{eq:fit_low}
\end{align}
Algorithm \ref{alg:low_pso} summarizes the proposed LBO-PSO algorithm, where the fitness function evaluation using (\ref{eq:fit_low}) at each iteration does not require channel samples and associated fitness evaluations (as done in Algorithm \ref{alg:batch_pso}).
With only the S-CSI and random channel distribution of the links, each particle $\boldsymbol{\theta}$ determines the optimal power solution using (\ref{eq:opt_pow}), and then evaluates its fitness value using (\ref{eq:fit_low}) at each iteration.
The best position of each particle and the global optimal position are updated according to the historical fitness values.
The velocities and positions of the particles are also updated using (\ref{eq:speed}) and (\ref{eq:pos}), respectively.
Upon the completion of the LBO-PSO algorithm, the passive beamforming matrix $\mathbf{\Theta}$ is optimized and the IRS is configured accordingly for the frame. At each time slot of the frame, the UE measures and reports the effective channel.
Based on the reported effective channels, the BS runs the water-filling algorithm to allocate the transmit powers for the data streams in the next time slot, as described in Sec. III-A.

\begin{algorithm}[t]
	\small
	\caption{Proposed LBO-PSO Algorithm}
	\label{alg:low_pso}
	\LinesNumbered
	\SetKwBlock{Begin}{Step 1:}{end}
	\Begin(\textbf{Swarm initialization}){
		\SetKwInOut{KwIn}{\textbf{Initialize the PSO parameters}} \KwIn{$\{P, w, c_1, c_2, \varepsilon_1, \varepsilon_2, I_{\mathrm{iter}}\}$.}
		\SetKwInOut{KwIn}{\textbf{Initialize the position and velocity}} \KwIn{$\{\boldsymbol{\theta}_p^{(0)}\}_{p=1}^{P},  \{\mathbf{v}_{p}^{(0)}\}_{p=1}^{P}$.}
		For each particle $\boldsymbol{\theta}_p^{(0)}$, evaluate the fitness value of each particle using (\ref{eq:fit_low}), and obtain the initial best position set $\{\dot{\boldsymbol{\theta}}_p\}_{p=1}^{P}$\;
		Find the global optimal position $\ddot{\boldsymbol{\theta}} = \arg\min_{\dot{\boldsymbol{\theta}}} \{ J^{(0)} (\dot{\boldsymbol{\theta}}_1), J^{(0)} (\dot{\boldsymbol{\theta}}_2), \cdots, J^{(0)} (\dot{\boldsymbol{\theta}}_P) \}$\;}
	\SetKwBlock{Begin}{Step 2:}{end}
	\Begin(\textbf{Iterative search}){
		\For{$i=1: I_{\mathrm{iter}}$}{{
				Update particle velocity $\{\mathbf{v}_{p}^{(i)}\}_{p=1}^{P}$ and position $\{\boldsymbol{\theta}_p^{(i)}\}_{p=1}^{P}$ using (\ref{eq:speed}) and (\ref{eq:pos})\;
				\For{$p=1: P$}{
					Given particle $\boldsymbol{\theta}_p^{(i)}$, evaluate the fitness value of particle $\boldsymbol{\theta}_p^{(i)}$ using (\ref{eq:fit_low})\;
					\eIf{$J^{(i)} < J^{(i-1)}$}
					{$\dot{\boldsymbol{\theta}}_p = \boldsymbol{\theta}_p^{(i)}$\;}{$\dot{\boldsymbol{\theta}}_p = \boldsymbol{\theta}_p^{(i-1)}$\;}
					Update $\ddot{\boldsymbol{\theta}}_p = \arg\min_{\dot{\boldsymbol{\theta}}} \{ J^{(i)} (\dot{\boldsymbol{\theta}}_1), J^{(i)} (\dot{\boldsymbol{\theta}}_2), \cdots, J^{(i)} (\dot{\boldsymbol{\theta}}_P) \}$\;
				}
			}
		}
	}
\end{algorithm}

It is important to note that the objective of Problem (P6) is an upper bound for the objective of Problem (P5).
In other words, the solution to Problem (P6) provides a lower bound for the AAR achievable by Problem (P1').
The lower bound is derived by first exploiting the arithmetic-geometric inequality~\cite{7166320}, followed by the Jensen's inequality, to reveal that the lower bound only depends on the trace of inverse channel covariance matrix. By applying the property of complex inverse Wishart matrix, we show that the expectation of the inverse channel covariance matrix only depends on the S-CSI and so does the AAR lower bound.
The tightness of the lower bound depends on the curvature of the AAR regarding the antenna number, the IRS element number, and the channel rank (due to the use of the Jensen's inequality), and is hard to quantify.
On the other hand, the AAR lower bound and subsequently Problem (P6) only require the S-CSI knowledge, i.e., the mean and covariance of the CSI, and do not need to generate channel samples (as opposed to the proposed mbs-PSO algorithm and more generic PSO-based algorithms). To this end, the LBO-PSO algorithm developed to solve Problem (P6) is immune to the uncertainty in the generation of channel samples and the subsequent potential inaccuracy of using the channel samples to approximate the real channel. Moreover, the LBO-PSO algorithm is computationally more efficient, due to the dramatically lower complexity for evaluating the PSO fitness value, as compared to the mbs-PSO. For these reasons, the LBO-PSO is a computationally efficient alternative to the mbs-PSO and provides assured results for qualifying the mbs-PSO's results.

\subsection{Complexity Analysis}
In what follows, we analyze the computational complexity of the two proposed PSO algorithms measured by floating point operations per second (FLOPS).

\textbf{mbs-PSO algorithm:} According to (\ref{eq:sinr}), the operations dominating the computational complexity of each particle include a matrix addition, four matrix multiplications, and a matrix inversion per channel sample per iteration, requiring $F_1 = 4 \left(N_{\mathrm{t}}+N_{\mathrm{r}} \right)N^2 +  4M N_{\mathrm{t}} N_{\mathrm{r}} + 4M^2  N_{\mathrm{r}} + (4M^3+M^2+M)$ FLOPS \cite{4917823}.
The overall computational complexity of the algorithm is $P \times L_{\mathrm{mb}} \times F_1$ FLOPS, according to (\ref{eq:fitness}). In contrast, a direct use of all $L_B$ channel samples to evaluate the fitness function per iteration using (\ref{eq:AAR_old}) incurs $P \times L_{B} \times F_1$ FLOPS, which is $N_B$ times the complexity of the new mbs-PSO algorithm since $L_{B}=N_B L_{\mathrm{mb}}$.

\textbf{LBO-PSO algorithm:} According to (\ref{eq:lowercom}), the operations dominating the complexity of each particle include three matrix additions, three matrix multiplications, a matrix inversion, and a matrix trace per iteration, incurring $F_2=4 \left(N_{\mathrm{t}}+N_{\mathrm{r}} \right)N^2 +  4 N_{\mathrm{t}}^2 N_{\mathrm{r}} +  (4N_{\mathrm{t}}^3+N_{\mathrm{t}}^2+N_{\mathrm{t}}) + 2N_{\mathrm{t}}^2$ FLOPS. Requiring no channel samples, the algorithm incurs $P \times F_2$ FLOPS.

In general, a moderate number of particles and iterations can lead to a reasonably good performance of the proposed Algorithms 1 and 2.
Moreover, the dominating computations of the fitness evaluation can be potentially parallelized for real-time implementation and pipelined in the field programmable gate array (FPGA) \cite{4064867, 1638140, huang2017pipeline}.

\begin{table*}[ht]
\centering\small
\caption{SIMULATION PARAMETERS}
\label{tb:params}
\begin{spacing}{1.2}
\begin{tabular}{|c|c||c|c|}
  \hline
  \textbf{Parameter} & \textbf{Value} & \textbf{Parameter} & \textbf{Value} \\
  \hline\hline
  Number of transmit antennas $N_{\mathrm{t}}$ & $8$          & Total transmit power $P_{\mathrm{tot}}$ & $20$ dBm \\
  \hline
  Number of receive antennas $N_{\mathrm{r}}$  & $4$          & Average noise power $\sigma_n^2$ & $-80$ dBm \\
  \hline
  Size of IRS meta-atoms $N_x \times N_y$      & $8\times 8$  & Number of channel samples $L_B$ &  $5000$ \\
  \hline
  Rician factor          $\kappa$              & $3$  & Number of channel samples per batch   & $50$ \\
  \hline
  Normalized Doppler frequency $\bar{f}_d$     & $0.01$  &  Swarm size $P$ & $200$ \\
  \hline
  PLE of the IRS-UE link $\alpha_{\mathrm{ur}}$ & $2.2$ & Number of PSO iterations $I_{\mathrm{iter}}$ & $100$ \\
  \hline
  PLE of the BS-IRS link $\alpha_{\mathrm{br}}$ & $2.2$ & Inertia weight of a particle $w$ & $0.9$ \\
  \hline
  PLE of the BS-UE link $\alpha_{\mathrm{bu}}$  & $3.6$ &  Cognitive scaling factor $c_1$ & $1.49445$ \\
  \hline
  Location of the BS                            & $[0, 0, 5]$     & Social scaling factor $c_2$  & $1.49445$ \\
  \hline
  Location of the IRS                           & $[100, 0, 5]$   & Reference path loss $L_{\mathrm{in}}$ &  $-30$ dB \\
  \hline
\end{tabular}
\end{spacing}
\end{table*}

\section{Simulation Results}\label{sec:sim}
Simulation results are provided to evaluate the performance of our proposed transmission method, including the transmission architecture and resource allocation algorithms.
Our testing environment is MATLAB R2018b on a desktop computer with 3.20 GHz Intel Core i7-8700 CPU and 8 GB random access memory (RAM).
In the simulations, the distance-dependent path loss model is given by $L_{\mathrm{out}}=L_{\mathrm{in}}(d/d_0)^{-\alpha}$ where $L_{\mathrm{in}}$ is the reference path loss at $d_0=1$ m and $\alpha$ is the path loss exponent (PLE).
Considering the 3D Cartesian coordinates, the BS and IRS are located on a three-dimensional plane with coordinates [0, 0, 5] m and [100, 0, 5] m.
The UE is 1 m high, distributed horizontally within the disk area with the center at [100, 10] m and radius of 10 m.
Unless otherwise stated, all other simulation parameters consisting of wireless propagation and PSO related parameters, are listed in Table \ref{tb:params}.

In addition to the proposed mbs-PSO and LBO-PSO schemes, five benchmark schemes are evaluated for comparison:
\begin{itemize}
\item \textbf{PSO with AAR fitness}: The PSO method is performed by using the objective of problem (P3), i.e., AAR, as the fitness function, where each iteration requires all generated channel samples to evaluate the fitness and update the particle positions. This method serves as an upper bound for the proposed mbs-PSO method.
\item \textbf{Sum-path-gain maximization (SPGM)-based scheme} \cite{9043523}: In our proposed two-timescale beamforming framework, the large-timescale passive beamforming optimization is replaced with the SPGM-based method, where the IRS reflection matrix is obtained by maximizing the effective channel gains.
\item \textbf{SVD-based scheme}: We adopt the conventional SVD precoder in the considered IRS-assisted MIMO system, where the BS optimizes its transmit powers solely based on the outdated CSI at each slot.
\item \textbf{Random phase shift}: Based on the studied SVD-ZF system, we adopt a randomly generated long-term reflection matrix, and then optimize the transmit powers by using the method developed in Sec. III-A.
\item \textbf{Without IRS}: We optimize the transmit powers to maximize the overall AAR of the studied SVD-ZF system with the IRS disabled.
\end{itemize}

\subsection{Convergence}

\begin{figure*}[t]
	\centering
    \subfigure[]{\includegraphics[scale=0.5]{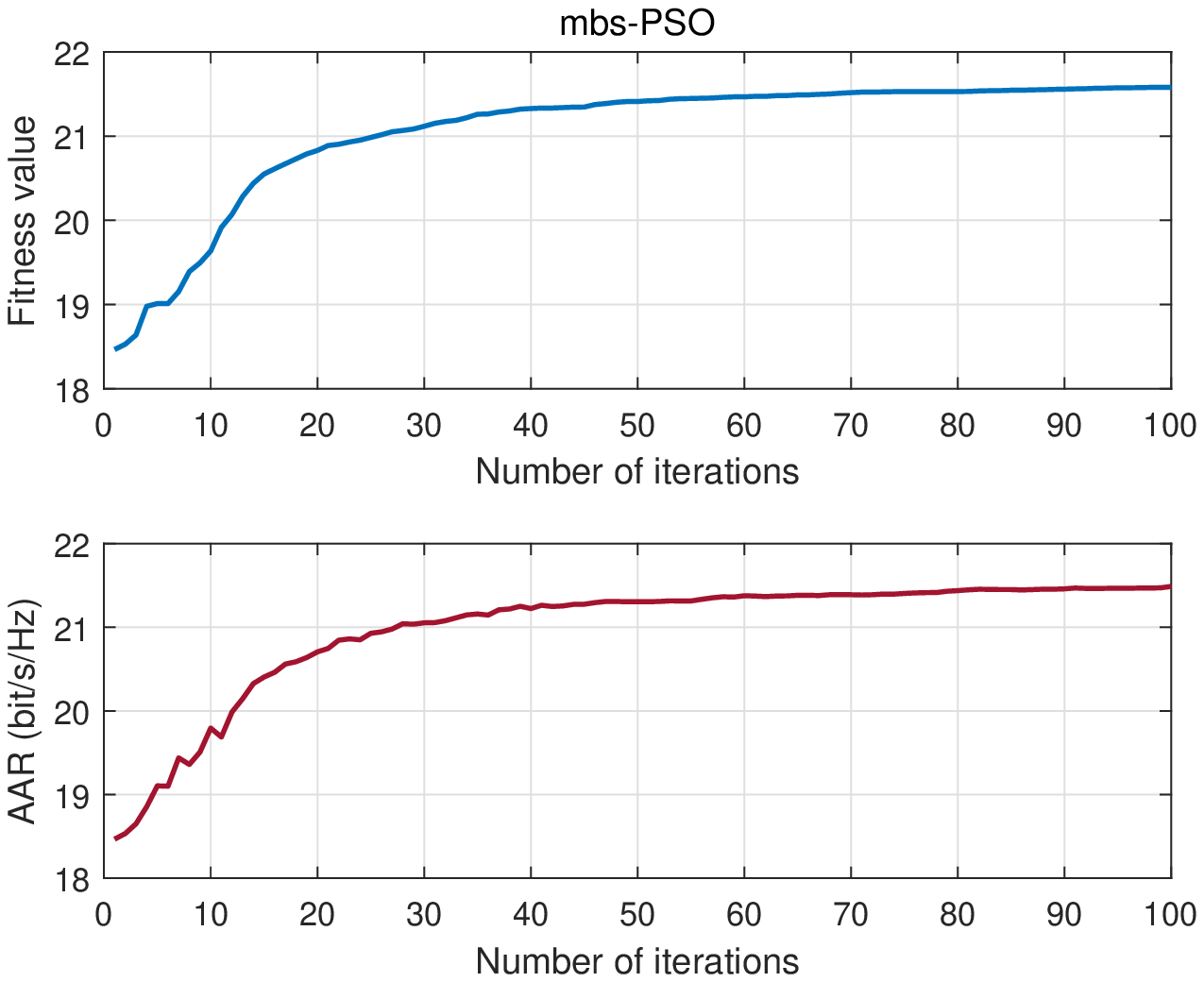}}
    \hspace{5mm}
    \subfigure[]{\includegraphics[scale=0.5]{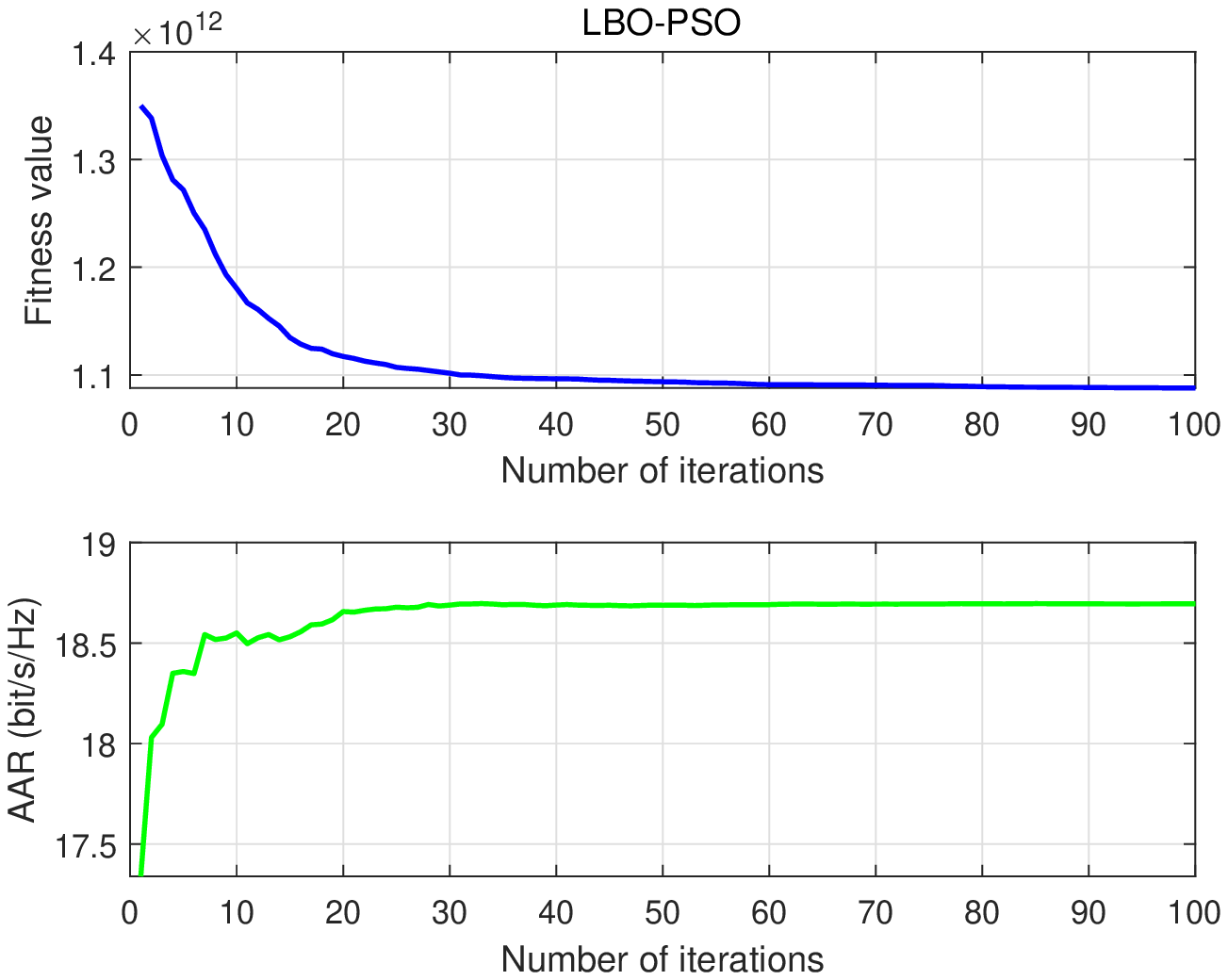}}
	\caption{Convergence of the proposed mbs-PSO and LBO-PSO, where the $x$-axis is the number of iterations, and the $y$-axis provides the fitness values of PSO and the AAR. }
	\label{fig:fig4}
\end{figure*}

Fig. \ref{fig:fig4} shows the convergence behaviors of the proposed mbs-PSO and LBO-PSO algorithms, where the fitness values of the two algorithms, defined in (\ref{eq:fitness}) and (\ref{eq:fit_low}), are separately plotted. From the fitness value results shown in Fig. \ref{fig:fig4}(a), we see that the mbs-PSO algorithm exhibits good convergence. More importantly, the mini-batch sampling surrogate function is effective, as the corresponding AAR significantly improves during iterations.
In Fig. \ref{fig:fig4}(b), the fitness function of the LBO-PSO algorithm is also shown to be effective.
It is worth noting that the fitness value plotted in Fig. 4(b) is the objective value of (P5), not the AAR in Problem (P1).
Problem (P5) is a TicMin problem. The fitness value measures the trace-of-inverse-covariance. Since the diagonal entries of the involved covariance matrix take relatively small values, the inverse of the covariance matrix (or in other words, the fitness value) is large.

\begin{figure}[t]
    \centering{}\includegraphics[scale=0.45]{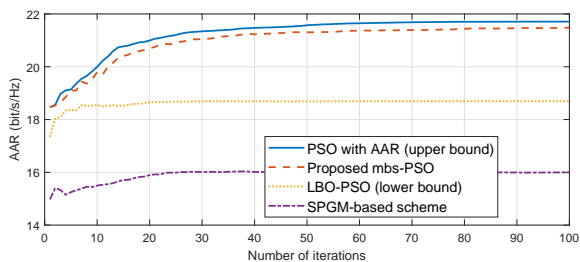}
	\caption{AAR achieved over iterations, with $N_{\mathrm{t}}=8$, $N_{\mathrm{r}}=4$, $N=64$ and $P_{\mathrm{tot}}=20$ dBm.}
	\label{fig:fig5}
\end{figure}

Fig. \ref{fig:fig5} shows the AAR of different schemes during PSO iterations in an $8 \times 4$ MIMO scenario. Since the PSO with AAR fitness uses a large number of channel samples to reasonably approximate the channel, it serves as an upper bound for the proposed mbs-PSO algorithm.
The LBO-based algorithm exploits only the S-CSI and serves as a lower bound for the mbs-PSO. As expected, the mbs-PSO scheme is between the lower bound and the upper bound, and considerably closer to the upper bound.

\begin{figure}[t]
    \centering{}\includegraphics[scale=0.35]{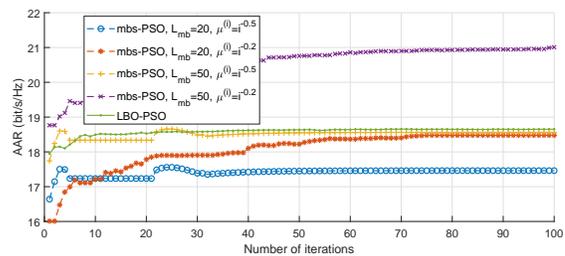}
	\caption{Impact of fitting parameters on the PSO performance.}
	\label{fig:fig6}
\end{figure}

Fig. \ref{fig:fig6} shows the impact of different fitting parameters, including mini-batch sizes and decay weight coefficients, on the convergent AAR. It is observed that the mbs-PSO algorithm can converge within 100 iterations under different $L_{\mathrm{mb}}$ and $\mu^{(i)}$ values.
The mbs-PSO algorithm can slightly improve the AAR over iterations under small $L_{\mathrm{mb}}$ values, while the improvement is substantial under large $L_{\mathrm{mb}}$ values. On the other hand, the AAR performance achieved by the mbs-PSO algorithm is better under the decay weight coefficient of $\mu^{(i)}=i^{-0.2}$ than it is under $\mu^{(i)}=i^{-0.5}$.
By comparing with the proposed LBO-PSO algorithm, we see that the mbs-PSO with small $L_{\mathrm{mb}}$ and small decay $\mu^{(i)}$ can deteriorate significantly. In particular, the LBO-PSO does not require regenerations of channel samples. It does not suffer from approximation inaccuracy from ill fitting parameters or dirty samples, and provides performance assurance and reliable benchmark for the proposed mbs-PSO algorithm.

\begin{figure}[t]
	\centering{}\includegraphics[scale=0.5]{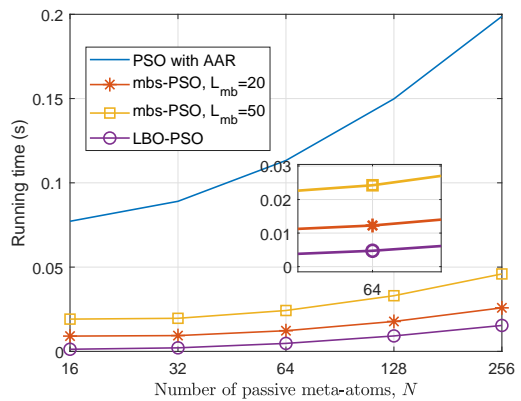}
	\caption{Running time per iteration vs. the number of IRS elements.}
	\label{fig:fig7}
\end{figure}

Fig. \ref{fig:fig7} shows that the running time of the proposed PSO algorithms is much shorter than that of the conventional PSO using full batch samples. Moreover, the processing speed of LBO-PSO is at a millisecond level when there are less than 64 IRS meta-atoms, and is considerably faster than the speed of the mbs-PSO algorithm.

\subsection{Impact of IRS}
\begin{figure}[t]
    \centering{}\includegraphics[scale=0.6]{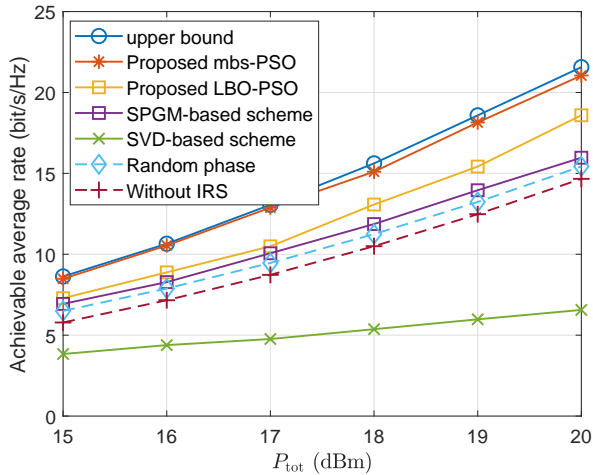}
	\caption{AAR performance vs. total transmit power under different schemes.}
	\label{fig:fig8}
\end{figure}

To demonstrate the usefulness of the IRS, Fig. \ref{fig:fig8} compares the different schemes under different transmit powers $P_{\mathrm{tot}}$. As $P_{\mathrm{tot}}$ increases, the AARs of the schemes improve with the SINR. Among the schemes, the conventional SVD scheme receives the worst performance, since it fails to suppress the inter-stream interference resulting from the outdated CSI. The proposed mbs-PSO algorithm, together with its upper bound, i.e., PSO with AAR fitness, and lower bound, i.e., the proposed LBS-PSO, surpass the other schemes, including ``random phase shift'' and ``without IRS'', confirming the usefulness of the IRS.
Furthermore, the SPGM-based scheme performs worse than the proposed algorithms. The reason is that the scheme optimizes the large-timescale passive beamforming by maximizing the effective channel gains, and does not utilize the channel correlation.
In contrast, our proposed algorithms optimize the large-timescale passive beamforming by considering the equivalent end-to-end channel (which is made up of the wireless channel and the precoder based on the oudated CSI), under the optimal small-timescale power allocation and ZF detection. In this sense, the channel correlation is better utilized in the proposed algorithms.

\begin{figure}[t]
    \centering{}\includegraphics[scale=0.6]{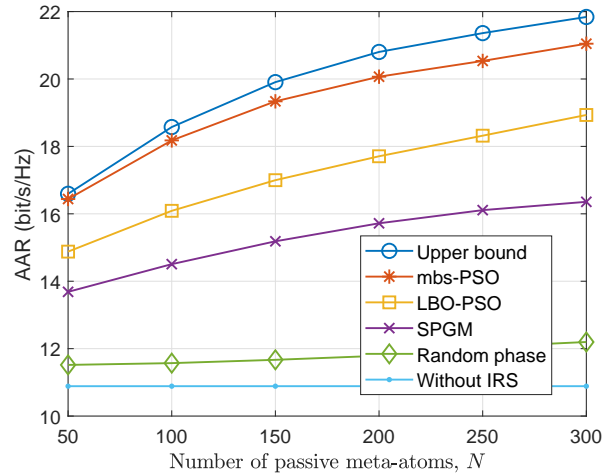}
	\caption{AAR performance vs. the number of reflecting elements.}
	\label{fig:fig9}
\end{figure}

\begin{figure}[t]
	\centering{}\includegraphics[scale=0.4]{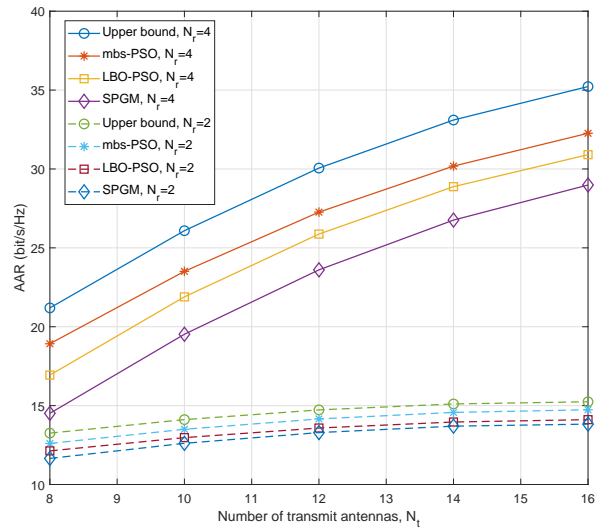}
	\caption{AAR performance vs. the number of transmit antennas.}
	\label{fig:fig10}
\end{figure}

Fig. \ref{fig:fig9} evaluates the impact of the reflecting element number on the system AAR. We see that all algorithms can improve their system AARs with the increasing number of IRS reflecting elements, except the scheme without IRS. This validates the effectiveness of the IRS and the proposed algorithm.
We also see that the AAR gap between the proposed mbs-PSO algorithm and the upper bound is small, demonstrating the validity of the mini-batch sampling method. As expected, the proposed LBO-PSO algorithm is lower than the proposed mbs-PSO algorithm in terms of the AAR. Nevertheless, their gap is consistent across a wide spectrum of the IRS element numbers.
Fig. \ref{fig:fig10} plots the AAR of each considered scheme with the increase of transmit antennas.
The AARs of all schemes grow with transmit antennas, but their growth rates slow down. Increasing the number of receive antennas is more effective than increasing the transmit antennas in terms of improving the AAR. Moreover, the SPGM-based method performs the worst among all methods for the reason revealed in Fig. \ref{fig:fig8}.

\subsection{Impact of Outdated CSI}
\begin{figure}[t]
    \centering{}\includegraphics[scale=0.4]{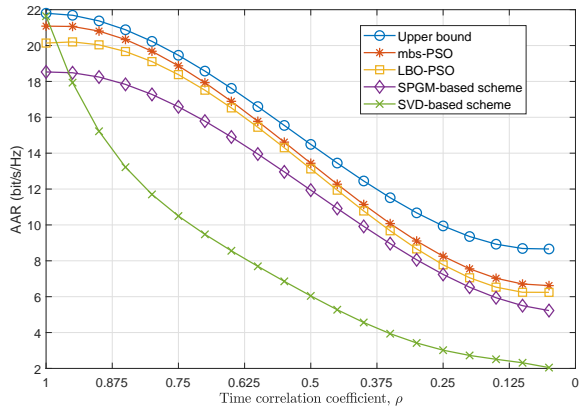}
	\caption{AAR vs. time correlation coefficient $\rho$.}
	\label{fig:fig11}
\end{figure}

Fig. \ref{fig:fig11} presents the AARs achieved by different schemes against the time correlation coefficient $\rho$.
A larger temporal correlation coefficient $\rho$ indicates a smaller time delay $\tau$, thus resulting in smaller channel error between consecutive slots.
In the extreme case where the outdated CSI is independent of the I-CSI, the precoding matrices of the schemes are independent of the I-CSI and equivalent to arbitrary precoding matrices.
We see that more severely outdated CSI leads to a lower AAR. This confirms the importance of exploiting the channel correlation between consecutive time slots to improve the AAR performance.
In the case of $\rho=1$, the CSI remains unchanged and never outdated throughout all time slots. It is seen that the conventional SVD scheme can achieve the same AAR as the upper bound, since SVD precoding is known to be optimal under the I-CSI.
Under the nearly independent channel conditions, i.e., $\rho \rightarrow 0$, the SVD precoder based on the outdated CSI is in essence an arbitrary precoder. Nevertheless, the proposed SVD-ZF-based scheme can still outperform the SVD-based scheme based on the outdated CSI, because of its use of the ZF detector to suppress inter-stream interference based on the I-CSI at the UE.

\section{Conclusion}\label{sec:con}
This paper developed a new SVD-ZF-based transmission strategy to overcome the undesired outdated CSI in IRS-assisted MIMO systems. Specifically, we proposed a two-timescale beamforming protocol, where passive beamforming at the IRS is designed at a large timescale and power allocation at the BS is performed slot by slot based on the outdated CSI at a small timescale.
The power allocation problem was derived with the closed-form water-filling solutions. The large-timescale passive beamforming problem was first transformed to a non-convex deterministic AAR maximization problem. We proposed an mbs-PSO algorithm to efficiently solve this non-convex problem. We also provided an LBO-PSO algorithm to obtain the passive beamformers with only S-CSI, and its resulting AAR was utilized as a lower bound for validation of the mbs-PSO algorithm. Finally, we evaluated the proposed two-timescale MIMO transmission strategy and PSO algorithms by comparing with multiple benchmarks, and concluded that the proposed mbs-PSO algorithm can approach the upper bound of the system AAR, and outperforms the existing benchmarks.

\begin{appendices}
\section{Proof of Lemma 1}\label{prf1}
Let $\mathbf{v}_i$ be a basis vector of $\mathbf{B}$ with $\mathbf{B} \mathbf{v}_i = \lambda_i \mathbf{v}_i$. Since $\mathbf{B}$ is symmetric, $\mathbf{v}_i^{\mathsf{H}} \mathbf{B} = \lambda_i \mathbf{v}_i^{\mathsf{H}}$ and
\begin{align}
	\mathrm{trace}(\mathbf{B}\mathbf{A}) &= \sum_{i=1}^N \mathbf{v}_i^{\mathsf{H}} \mathbf{B}\mathbf{A} \mathbf{v}_i = \sum_{i=1}^N \lambda_i \mathbf{v}_i^{\mathsf{H}} \mathbf{A} \mathbf{v}_i \nonumber\\
	&\leq \sum_{i=1}^N \lambda_{\max}(\mathbf{B})  \mathbf{v}_i^{\mathsf{H}} \mathbf{A} \mathbf{v}_i = \lambda_{\max}(\mathbf{B}) \sum_{i=1}^N \mathbf{v}_i^{\mathsf{H}} \mathbf{A} \mathbf{v}_i \nonumber\\
	&= \lambda_{\max}(\mathbf{B}) \mathrm{trace}(\mathbf{A}).
\end{align}
This proof completes. \QEDA
\end{appendices}


\begin{IEEEbiography}[{\includegraphics[width=1in,height=1.25in,clip,keepaspectratio]{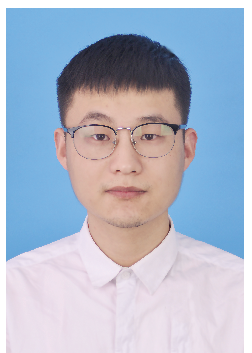}}]{Yashuai Cao} (S'18) received the B.S. degree from Chongqing University of Posts and Telecommunications (CQUPT), Chongqing, China, in 2017. He is currently pursuing the Ph.D. degree in communication engineering with the School of Information and Communication Engineering, Beijing University of Posts and Telecommunications (BUPT), Beijing, China. His current research interests include wireless resource allocation and signal processing technologies for massive MIMO systems and intelligent reflecting surface assisted wireless networks.
\end{IEEEbiography}

\begin{IEEEbiography}[{\includegraphics[width=1in,height=1.25in,clip,keepaspectratio]{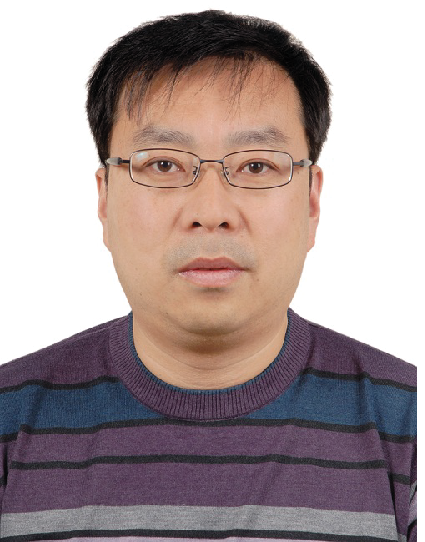}}]{Tiejun Lv}
(M'08-SM'12) received the M.S. and Ph.D. degrees in electronic engineering from the University of Electronic Science and Technology of China (UESTC), Chengdu, China, in 1997 and 2000, respectively. From January 2001 to January 2003, he was a Postdoctoral Fellow with Tsinghua University, Beijing, China. In 2005, he was promoted to a Full Professor with the School of Information and Communication Engineering, Beijing University of Posts and Telecommunications (BUPT). From September 2008 to March 2009, he was a Visiting Professor with the Department of Electrical Engineering, Stanford University, Stanford, CA, USA. He is the author of three books, more than 100 published IEEE journal papers and 200 conference papers on the physical layer of wireless mobile communications. His current research interests include signal processing, communications theory and networking. He was the recipient of the Program for New Century Excellent Talents in University Award from the Ministry of Education, China, in 2006. He received the Nature Science Award in the Ministry of Education of China for the hierarchical cooperative communication theory and technologies in 2015.
\end{IEEEbiography}

\begin{IEEEbiography}[{\includegraphics[width=1in,height =1.25in,clip,keepaspectratio]{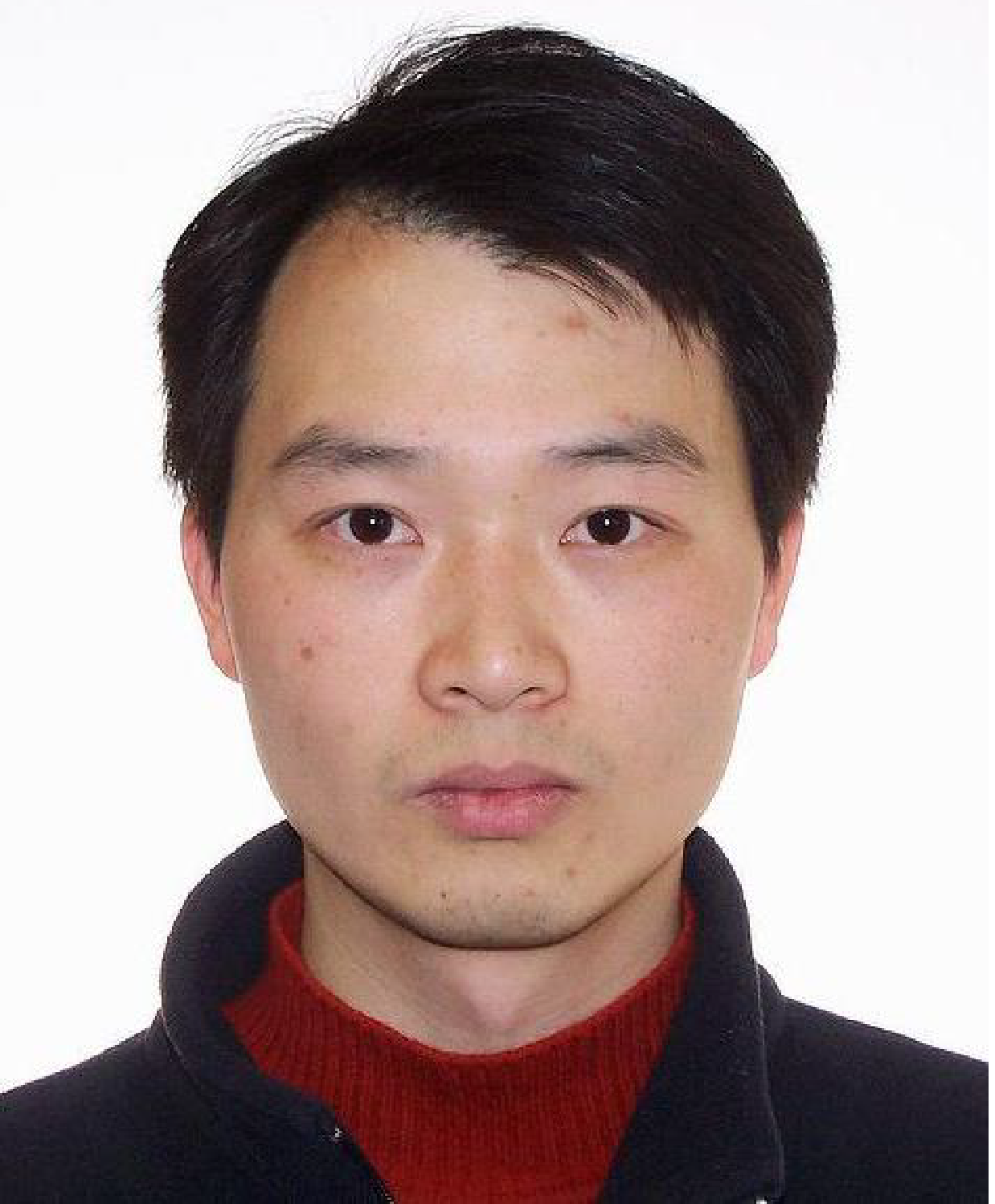}}]{Wei Ni} (M'09-SM'15) received the B.E. and Ph.D. degrees in Electronic Engineering from Fudan University, Shanghai, China, in 2000 and 2005, respectively. Currently, he is a Principal Research Scientist at CSIRO, Sydney, Australia, and an Adjunct Professor at the University of Technology Sydney and Honorary Professor at Macquarie University. He was a Postdoctoral Research Fellow at Shanghai Jiaotong University from 2005 to 2008; Deputy Project Manager at the Bell Labs, Alcatel/Alcatel-Lucent from 2005 to 2008; and Senior Researcher at Devices R\&D, Nokia from 2008 to 2009. He has authored five book chapters, more than 200 journal papers, more than 80 conference papers, 25 patents, and ten standard proposals accepted by IEEE. His research interests include machine learning, online learning, stochastic optimization, as well as their applications to system efficiency and integrity.
Dr Ni is the Chair of IEEE Vehicular Technology Society (VTS) New South Wales (NSW) Chapter since 2020, an Editor of IEEE Transactions on Wireless Communications since 2018, and an Editor of IEEE Transactions on Vehicular Technology. He served first the Secretary and then the Vice-Chair of IEEE NSW VTS Chapter from 2015 to 2019, Track Chair for VTC-Spring 2017, Track Co-chair for IEEE VTC-Spring 2016, Publication Chair for BodyNet 2015, and Student Travel Grant Chair for WPMC 2014.
\end{IEEEbiography}

\end{document}